\documentclass[twocolumn,aps]{revtex4}
\usepackage{times}
\usepackage{natbib}
\usepackage{graphicx}
\usepackage{amssymb,amsmath,amsbsy}

\begin{document}
\title{Characterization of single photons using two-photon interference}
\author{T. Legero}
\altaffiliation[Current address: ]{Physikalisch-Technische Bundesanstalt, Bundesallee 100, 38116 Braunschweig, Germany}
\author{T. Wilk}
\author{A. Kuhn}
\author{G. Rempe}
\email[Corresponding author: ]{gerhard.rempe@mpq.mpg.de}
\affiliation{Max-Planck-Institut f\"{u}r Quantenoptik,\\ Hans-Kopfermann-Str. 1, 85748 Garching, Germany}
\maketitle
\tableofcontents
\vspace{5mm}

\noindent{\small It is a pleasure for us to dedicate this paper to Prof. Herbert Walther, a pioneer in quantum optics from the very beginning. The investigation of the amazing properties of single photons both in the microwave and the optical domain has always been a central theme in his research. We wish him all the best in the years to come!}

\section{Introduction}
Four decades after the pioneering work on optical coherence and photon statistics by Glauber \cite{Glauber65}, the controlled generation of single photons with well-defined coherence properties is now of fundamental interest for many applications in quantum information science. First, single photons are an important ingredient for quantum cryptography and secure quantum key distribution  \citep{Gisin02}. Second, the realization of quantum computing with linear optics (LOQC), which was first proposed by Knill, Laflamme and Milburn \cite{Knill01}, relies on the availability of deterministic single-photon sources. And third, various schemes have been proposed to entangle and teleport the spin of distant atoms, acting as emitters of single photons, by means of correlation measurements performed on the single-photon light fields \citep{Cabrillo99, Bose99, Browne03, Duan03:2}. Therefore, in recent years, a lot of effort has been made to realize single-photon sources. As a result, the controlled generation of single photons has been demonstrated in various systems, as summarized in a review article of Oxborrow and Sinclair \cite{Oxborrow05}.

Using the process of spontaneous emission from a single quantum system is the simplest way to realize a single-photon source. In this case, the quantum system is excited by a short laser pulse and the subsequent spontaneous decay of the system leads to the emission of only one single photon. This has been successfully demonstrated many times, e.g., using single molecules \citep{Brunel99, Lounis00, Moerner04}, single atoms \citep{Darquie05}, single ions \citep{Blinov04}, single color centers \citep{Kurtsiefer00, Brouri00, Gaebel04} or single semiconductor quantum dots \citep{Santori01, Yuan02, Pelton02, Aichele04}. If the quantum system radiates into a free-space environment, the direction of the emitted photon is unknown. This limits the efficiency of the source. To overcome this problem, the enhanced spontaneous emission into a cavity has been used. The system is coupled to a high-finesse cavity and the photon is preferably emitted into the cavity mode, which defines the direction of the photons. Although a cavity is used, most properties of the photons, like the frequency, the duration and the bandwidth, are given by the specific quantum system. Only if the generation of single photons is driven by an adiabatic passage can these spectral parameters be controlled. This technique uses the atom-cavity coupling and a laser pulse to perform a vacuum stimulated Raman-transition (STIRAP), which leads to the generation of one single photon. Up to now, this has been demonstrated with single Rubidium atoms \citep{Hennrich04,Kuhn02:2}, single Caesium atoms \citep{McKeever04} and single Calcium ions \citep{Keller04} placed in high-finesse optical cavities.

The characterization of a single-photon source usually starts with the investigation of the photons statistics, which is done by a $g^{(2)}$ correlation measurement using a Hanbury Brown and Twiss \cite{HanburyBrown57} setup. The observation of antibunching shows that the source emits only single photons. However, the requirements on a single-photon source for LOQC and for the entanglement of two distant atoms go far beyond the simple fact of antibunching. The realization of these proposals relies on the indistinguishability of the photons, so that even photons from
different sources need to be identical with respect to their frequency, duration and shape. Therefore it is desirable to investigate the spectral and temporal properties of single photons emitted from a given source. We emphasize that properties like bandwidth or duration always deal with an ensemble of photons and cannot be determined from a measurement on just a single photon. Therefore any measurement of these properties requires a large ensemble of successively emitted photons. Several methods have been employed to characterize these.

The first measurement of the duration of single photons has been performed by Hong, Ou and Mandel \cite{Hong87}. In this experiment, the fourth-order interference of two photons from a parametric down-conversion source was investigated by superimposing the signal and the idler photon on a 50/50 beam splitter. The coincidence rate of photodetections at the two output ports of the beam splitter was measured in dependence of a relative arrival-time delay between the two photons. Indistinguishable photons always leave the beam splitter together, so that no coincidence counts can be observed. If the photons are slightly different, e.g. because they impinge on the beam splitter at slightly different times, the coincidence rate increases. Therefore, as a function of the photon delay, the coincidence rate shows a minimum if the photons impinge simultaneously on the beam splitter, and for otherwise identical photons the width of this dip is the photon duration. The minimum in the coincidence rate goes down to zero if the photons are identical. Any difference between the two interfering photons reduces the depth of this dip. The first demonstration of such a two-photon interference of two independently emitted photons from a quantum-dot device has been shown by Santori et al. \cite{Santori02}.

In addition to the two-photon coincidence experiments, a correlation measurement between the trigger event and the detection time of the generated photon can be used to determine the temporal envelope of the photon ensemble \citep{Kuhn02:2, McKeever04,Keller04}. This latter method is insensitive to the spectral properties of the photons. In general, it does not reveal the shape of the single-photon wavepackets, unless all photons are identical. In case of variations in the photon duration or a jitter in the emission time, only the temporal envelope of the photon ensemble is observed. No conclusions can be drawn on the envelope of the individual photons.

The standard way to determine the coherence time of a given light source is the measurement of the second-order interference using a Mach-Zehnder or Michelson interferometer. This measurement can also be done with single photons, so that each single photon follows both paths of the interferometer and interferes with itself. The detection probability of the photons at both outputs of the interferometer shows a fringe pattern if the length of one arm is varied. The visibility of this pattern depends on the length difference of both arms and determines the coherence length (or the coherence time) of the photons. This method has been used by Santori et al. \cite{Santori02} and Jelezko et al. \cite{Jelezko03} to measure the coherence time of their single-photon sources. However, this method is hardly feasible for photons of long duration, because the length of one arm of the interferometer must be varied over large distances. Furthermore, the measurement depends on the mechanical stability of the whole setup, i.e. the interferometer must be stable within a few per cent of the wavelength of the photons, which might not be easy.

Only recently, adiabatic passage techniques have allowed the generation of photons which are very long compared to the detector time resolution. Therefore the detection time of a photon can be measured within the duration of the single-photon wavepacket. As a consequence, the two-photon interference can be investigated in a time-resolved manner, i.e., the coincidence rate can be measured as a function of the time between photodetections \citep{Legero03,Legero04}. The theoretical analysis shows that this method not only gives information about the duration of single photons, but also about their coherence time. Here we discuss how to use this method for a spectral or temporal characterization of a single-photon source.

The article is organized as follows: After a brief summary of the nature of single-photon light fields (Chapter \ref{sec:light_fields}), we discuss the interference of two photons on a beam splitter and introduce the time-resolved two-photon interference (Chapter \ref{sec:Two-photon_interference}). Thereafter, we show how a frequency and an emission-time jitter affects the results of a time-resolved two-photon interference experiment (Chapter \ref{sec:Jitter}). On this basis, the experimental characterization of a single-photon source, based on an adiabatic passage technique, is discussed (Chapter \ref{sec:Experiment}).

\section{Single-photon light fields}\label{sec:light_fields}
The quantum theoretical description of light within an optical cavity is well understood \citep{Meystre98}. The electromagnetic field between the two mirrors is subject to boundary conditions which lead to a discrete mode structure of the field. Each mode can be labeled by a number $l$ and is characterized by its individual frequency, $\omega$. These eigenfrequencies are separated by $\Delta \omega = 2 \pi c /L$, where $L$ is the round-trip length of the cavity. The quantization results in a discrete set of energies, $E_n = \hbar \omega (n + 1/2)$ and the appropriate eigenstates are defined by means of creation, $\hat{a}^\dag_l$, and annihilation, $\hat{a}^{}_l$, operators. The
energy eigenstates
\begin{equation}
|n\rangle = \frac{(\hat{a}^\dag_l)^n}{\sqrt{n!}} |0\rangle
\end{equation}
are states with a fixed photon number, $n$. In this context, photons are the quanta of energy in the modes of the cavity.

In the limit of $L \rightarrow \infty$ and $\Delta \omega \rightarrow 0$, the mode spectrum becomes continuous. In this case it is convenient to introduce continuous-mode operators $\hat{a}^\dag(\omega)$ and $\hat{a}(\omega)$ according to
\begin{equation}
\hat{a}^\dag_l \rightarrow (\Delta
\omega)^{1/2}\hat{a}^\dag(\omega)
\end{equation}
These operators create and annihilate photons as quanta of monochromatic waves in free space. These waves of infinite spatial extension do not have any beginning or any end. However, photons generated in a laboratory are characterized by a certain frequency bandwidth or a finite spatial extension. Therefore it is desirable to define operators which create or annihilate photons in modes of a given bandwidth, or, in other words, of a well defined spatiotemporal spread.

\subsection{Frequency modes}\label{sec:frequency_modes}
In contrast to modes describing monochromatic waves, it is possible to define field modes of a given frequency distribution. These modes represent wavepackets travelling with the speed of light $c$ through the vacuum, and the bandwidth $\kappa$ of such a mode determines the duration $\delta t$ of the wavepacket. A frequency distribution is described by a normalized complex function $\chi(\omega)$ which is called the mode function of the field. The operators $\hat{a}^\dag(\omega)$ and $\hat{a}(\omega)$ can be used to define a new set of operators for the creation and annihilation of photons in these new modes \citep{Blow90}. The creation operator, e.g., is given by
\begin{equation}\label{eq:Paket_Erzeuger_omega}
\hat{b}^\dagger_\chi = \int d \omega ~ \chi(\omega) ~
\hat{a}^\dagger(\omega).
\end{equation}
Note that the mode function $\chi(\omega)$ can be written as the product of a real amplitude, $\varepsilon(\omega)$, and a complex phase, $\exp{(-i \Phi(\omega))}$. The phase term includes the emission time $\tau_0$ and the propagation of the wavepacket. In the following, we restrict our discussion to Gaussian wavepackets centered at the frequency $\omega_0$. Their mode functions read
\begin{equation}\label{eq:Gausssche-Frequenzverteilung}
\chi(\omega) = \sqrt[4]{\frac{2}{\pi \kappa^2}} ~
\exp{\left(-\frac{(\omega - \omega_{0})^2}{\kappa^2}\right)}
\exp{(-i \omega (\tau_0 + z/c))}.
\end{equation}
For an ideal single-photon source which always produces identical photons, the light field is always described by the same quantum mechanical state vector. In other words, the state vector is given by the creation operator $\hat{b}^\dagger(\chi)$ acting on the vacuum state $|0\rangle$ for every single photon:
\begin{equation}
|1_\chi\rangle^{} = \hat{b}^\dagger_\chi ~ |0\rangle.
\end{equation}
We emphasize that such an ideal source is hardly feasible. Usually the generation process cannot be controlled perfectly and therefore the mode function is subject to small variations. To take this into account the light field must be described by a quantum mechanical density operator
\begin{equation}\label{eq:Dichtematrix}
\hat{\varrho} = \int \mathrm{d}\vartheta ~ f(\vartheta) ~
|1_{\chi(\vartheta)}\rangle\langle1_{\chi(\vartheta)}|.
\end{equation}
Here we assume that the source produces single photons with a Gaussian frequency distribution and the parameters of this distribution are subject to small variations, according to a distribution function $f(\vartheta)$. The parameter $\vartheta$ stands for the center frequency, $\omega_0$, the bandwidth, $\kappa$, or the emission time, $\tau_0$, of the photon, or a combination of these.

\subsection{Spatiotemporal modes}
Due to the Fourier theorem, each mode with a certain frequency distribution $\chi(\omega)$ can be assigned to a temporal wavepacket which is travelling through space. A mode with the Gaussian frequency distribution given by Eq.\,(\ref{eq:Gausssche-Frequenzverteilung}) therefore belongs to a spatiotemporal mode $\xi(t - z/c)$ of Gaussian shape. With the substitution $q:=t-z/c$ this mode is given by the function
\begin{equation}\label{eq:Gausssche-Raum-Zeit-Verteilung}
\begin{array}{rcl}
\xi(q) &=& \sqrt[4]{\frac{2}{\pi \delta t^2}} ~
\exp{\left(-\frac{q^2}{\delta t^2}\right)} \exp{(i \omega_0(\tau_0 - q))}\\[3mm]
&\equiv&\epsilon(q)~
\exp{(i \omega_0(\tau_0 - q))}.
\end{array}
\end{equation}
The duration $\delta t$ of this Gaussian wavepacket is given by the reciprocal bandwidth of the frequency distribution, $\delta t = 2/\kappa$. Blow et al. \cite{Blow90} have shown that creation and annihilation operators can also be assigned to spatiotemporal modes. In order to do that, one has to define the Fourier-transformed operators
\begin{eqnarray}
\hat{a}^\dagger(q) &=& (2\pi)^{-1/2} \int d\omega ~
\hat{a}^\dagger(\omega)
~ e^{-i\omega q}, \label{Fourier_transformed_operators_1} \\
\hat{a}^{}(q) &=& (2\pi)^{-1/2} \int d\omega ~ \hat{a}^{}(\omega)
~ e^{i \omega q}. \label{Fourier_transformed_operators_2}
\end{eqnarray}
By means of these operators we define a flux operator $\hat{a}^\dagger(q) \hat{a}^{}(q)$. Its expectation value has the unit of photons per unit time. We need this operator in the next section to describe the detection of single photons.

Eqs.\,(\ref{Fourier_transformed_operators_1}) and (\ref{Fourier_transformed_operators_2}) are only valid if the bandwidth of the modes is much smaller than the frequency of the light, $\kappa << \omega_0$. This also limits the localization of a single photon in such a spatiotemporal mode. In case of optical frequencies, this condition is usually fulfilled.

In analogy to Eq.\,(\ref{eq:Paket_Erzeuger_omega}) the Fourier-transformed operators can be used to define creation and annihilation operators for photons of spatiotemporal modes $\xi(q)$:
\begin{equation}\label{eq:Paket_Erzeuger_zeit}
\hat{c}^\dagger_\xi = \int dq ~ \xi(q) ~ \hat{a}^\dagger(q)
\end{equation}
To take fluctuations into account, one can again write the density operator of the light field as in Eq.\,(\ref{eq:Dichtematrix}), but using spatiotemporal modes. In this case, $\vartheta$ stands for any combination of $\omega_0, \delta t$, and $\tau_0$.

\subsection{Single-photon detection}\label{sec:detect_probability}
Choosing spatiotemporal modes for describing the state of a single-photon light field simplifies the formal description of the detection of a photon. We assume a detector with quantum efficiency $\eta$ placed at the position $z=0$. The response of the detector within a time interval $[t_0 - \mathrm{d}t_0/2, t_0 + \mathrm{d}t_0/2]$ is given by the expectation value of the flux operator:
\begin{equation}\label{eq:detection_probability}
P^{(1)}(t_0) = \eta ~ \int^{t_0 - \mathrm{d}t_0/2}_{t_0 +
\mathrm{d}t_0/2} \!\! \mathrm{d}t ~ \mathrm{tr}[~\hat{\varrho}^{}
~ \hat{a}^\dagger(t) \hat{a}^{}(t)~]
\end{equation}
In case of single-photon wavepackets, the function $P^{(1)}(t_0)$ gives the probability to detect this photon within the considered time interval. In practice, the lower limit of the duration $\mathrm{d}t_0$ is given by the detector time resolution $T$, i.e., $\mathrm{d}t_0 \geq T$.

If the photon duration is much longer than the detector time resolution, $\delta t >> T$ and $dt_0 =T$, Eq.\,(\ref{eq:detection_probability}) can be simplified to
\begin{equation}\label{eq:detection_probability_timeresolved}
P^{(1)}(t_0) = \eta ~T ~ \mathrm{tr}[~\hat{\varrho}^{} ~
\hat{a}^\dagger(t_0) \hat{a}^{}(t_0)~].
\end{equation}
The measurement of the detection probability requires a  large ensemble of single photons. In the following, we therefore assume a periodic stream of single photons emitted one-after-the-other, so that the photons always hit the detector one by one. If all photons of this stream are identical, the light field can simply be described by a state vector $|1_\xi\rangle^{}$ and the density operator is given by $\hat{\varrho} = |1_\xi\rangle \langle 1_\xi |$, with $|1_\xi\rangle=\hat{c}^\dagger_\xi |0\rangle$. In this case, the average detection probability of the ensemble of photons is given by the square of the absolute value of the mode function, $\xi(q)$, and is therefore identical to the shape of each individual photonic wavepacket,
\begin{equation}\label{eq:detection_probability_pure}
P^{(1)}(t_0) = \eta T ~ |\xi(t_0)|^2 = \eta T ~ \epsilon^2(t_0).
\end{equation}
As already discussed, the photons may differ from one another, and the density operator is given according to Eq.\,(\ref{eq:Dichtematrix}). The average detection probability is then given by
\begin{equation}\label{eq:detection_probability_mixed}
P^{(1)}(t_0) = \eta T ~\int \mathrm{d}\vartheta ~ f(\vartheta) ~
\epsilon^2(t_0,\vartheta).
\end{equation}
To obtain this equation, we assume that trace and integration can be exchanged. Obviously the average detection probability for the photon ensemble differs from that for individual photons. The average detection probability is, in general, affected by the variation, $f(\vartheta)$, of the parameters of the mode function, $\xi(t)$. Therefore it shows only a temporal envelope of the photon ensemble. However, the effect of each parameter onto $P^{(1)}(t_0)$ can be very different. A variation of the frequency, e.g., does not affect the real amplitude of the mode function, so that the average detection probability, $P^{(1)}(t_0)$, is simply given by Eq.\,(\ref{eq:detection_probability_pure}). This is not the case for variations of the other parameters, as will be shown in Chapter \ref{sec:Jitter}.

\section{Two-photon interference}\label{sec:Two-photon_interference}
We now consider two independent streams of Gaussian-shaped single photons that impinge on a 50/50 beam splitter such that always two photons are superimposed. As we show in Chapter \ref{sec:Experiment}, these two streams can originate from one single-photon source by directing each photon randomly into two different paths of suitable length, so that successively generated photons hit the beam splitter at the same time. Here we ask for the probability to detect the photons of each pair in different output ports of the beam splitter. In case of identical photons, the joint detection probability is zero. With polarization-entangled  photon pairs emitted from a down-conversion source, this effect has first been used by Alley and Shih \cite{Alley86} to test the violation of Bell's inequality by joint photodetections, and one year later, Hong, Ou and Mandel \cite{Hong87} have used it to measure the delay between two photons with sub-picosecond precision. Recently, two-photon interference phenomenona have successfully been employed to test the indistinguishability of independently generated single photons \citep{Santori02}. To illustrate this interference effect, we first assume that each photon of a given stream can be described by the same quantum mechanical state vector, $|1_{\xi}\rangle$, but allow the state vectors of the two considered streams to differ from one another. In Chapter \ref{sec:Jitter} we generalize this discussion to streams of photons which show a variation in the parameters of the mode functions, e.g. a variation in the photon frequency. Finally we show that the interference of photon pairs reveals information about the variations of the mode functions.

In section \ref{beam_splitter} and \ref{principle} we start with a brief discussion of the beam splitter and the principle of the two-photon interference. Afterwards we analyze the joint detection probability for photons in the limits of a photon that is either very short or very long compared to the detector time resolution.

\subsection{Quantum description of the beam splitter}\label{beam_splitter}
\begin{figure}[!t]
\begin{center}
\includegraphics[angle=0, width=0.85\columnwidth]{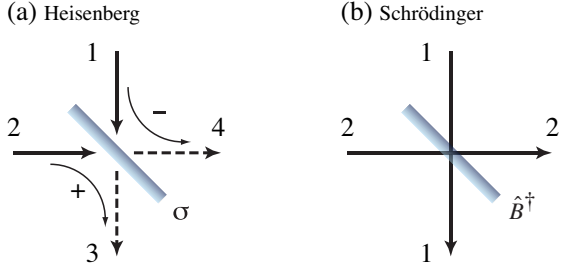}
\end{center}
\caption{The ideal lossless beam splitter is fully characterized by its transmission coefficient $\sigma$. The reflection coefficient is then given by $\sqrt{1-\sigma}$. Light can enter the beam splitter through two different input ports 1 and 2. According to the transmission and the reflection coefficient, it is divided into the output ports 3 and 4. For one of the reflections, the light is subject to a phase jump of $\pi$ which is indicated by the minus sign. In the Heisenberg picture \textbf{(a)}, one accounts for this process by transforming the creation and annihilation operators of the two input modes (1 and 2) into suitable operators of the output modes (3 and 4), whereas in the Schr\"{o}dinger picture \textbf{(b)}, the action of the beamsplitter is expressed by the unitary operator $\widehat{B}^\dagger$, which acts on the wavevector and couples the two through-going modes (1 and 2).}
\label{fig:Strahlteiler}
\end{figure}

The beam splitter is an optical four-port device with two inputs and two outputs. The principle of the beam splitter is shown in Figure\,\ref{fig:Strahlteiler}. As discussed by Leonhardt \cite{Leonhardt97}, each port has its own creation and annihilation operators, and the output operators can be expressed by the input operators using a unitary transformation matrix $\mathbf{B}$. This relation is valid for creation and annihilation operators $\hat{a}(\omega)$ of monochromatic waves as well as for operators of spatiotemporal modes, $\hat{b}_\chi$ or $\hat{c}_\xi$. It reads:
\begin{equation}\label{eq:QM_BS_Matrix}
\begin{pmatrix}
  \hat{a}^{}_{3} \\ \hat{a}^{}_{4}
\end{pmatrix} =
  \mathbf{B}
\begin{pmatrix}
  \hat{a}^{}_{1} \\ \hat{a}^{}_{2}
\end{pmatrix}
\quad \mathrm{and} \quad \left(\hat{a}^{\dagger}_{3},
\hat{a}^{\dagger}_{4}\right) = \left(\hat{a}^{\dagger}_{1},
\hat{a}^{\dagger}_{2}\right) \mathbf{B^*}
\end{equation}
In the following discussion, we assume an ideal lossless and polarization independent beam splitter with transmission coefficient $\sqrt{\sigma}$. The matrix of this beam splitter is given by
\begin{equation}\label{eq:einfacher_BS}
\mathbf{B} = \begin{pmatrix}
 \sqrt{\sigma}  & \sqrt{1-\sigma}  \\
  -\sqrt{1-\sigma} & \sqrt{\sigma}
\end{pmatrix}.
\end{equation}
The opposite signs of the off-diagonal terms reflect the phase jump of $\pi$ for the reflection at one side of the beam splitter.

The transmission of photons from the input side to the output side of the beam splitter can be understood as a quantum mechanical evolution of the system. This evolution can be described in two equivalent pictures corresponding to the Heisenberg and the Schr\"{o}dinger picture in quantum mechanics \citep{Campos89, Leonhardt03}. In the Heisenberg picture, the evolution is described by the creation and annihilation operators. The output operators are considered as the evolved input operators whereas the state vector of the field remains unchanged (see Figure \ref{fig:Strahlteiler} (a)). Using the unitary Operators $\widehat{B}$ and $\widehat{B}^\dagger$, this evolution can also be expressed by
\begin{equation}\label{eq:Entwicklung}
\begin{array}{c}
\begin{pmatrix}
  \hat{a}^{}_{3} \\ \hat{a}^{}_{4}
\end{pmatrix}  =
\mathbf{B}
\begin{pmatrix}
  \hat{a}^{}_{1} \\ \hat{a}^{}_{2}
\end{pmatrix} =:
  \widehat{B}
\begin{pmatrix}
  \hat{a}^{}_{1} \\ \hat{a}^{}_{2}
\end{pmatrix} \widehat{B}^\dagger\\[5mm]
\quad \mathrm{and} \quad \\[2mm]
\mathbf{B}^*
\begin{pmatrix}
  \hat{a}^{}_{1} \\ \hat{a}^{}_{2}
\end{pmatrix}
=:
\widehat{B}^\dagger
\begin{pmatrix}
  \hat{a}^{}_{1} \\ \hat{a}^{}_{2}
\end{pmatrix} \widehat{B}.
\end{array}
\end{equation}
Alternatively, in the Schr\"{o}dinger picture, the evolution can be calculated using the state vector of the light field. In this case, the state vector of the input side $|\Psi_{in}\rangle$ evolves to a state vector at the output side, $|\Psi_{out}\rangle = \widehat{B}^\dagger |\Psi_{in}\rangle$, while the modes themselves do not change, that is modes 1 and 2 are defined as the through-going modes (see Figure \ref{fig:Strahlteiler} (b)). In the next section the Schr\"{o}dinger picture is used to illustrate the principle of the two-photon interference.

\begin{figure}[t]
\begin{center}
\includegraphics[angle=0, width=1\columnwidth]{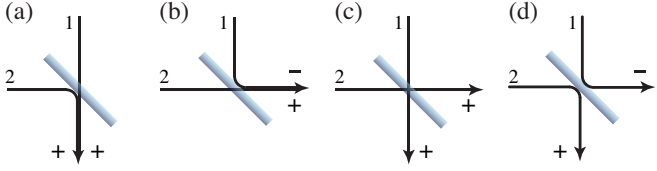}
\end{center}
\caption{Two impinging photons lead to four possible photon distributions at the beam-splitter output. In the first two cases (a) and (b) the photons would be found together. In the remaining two cases (c) and (d) the photons would leave the beam splitter through different ports. Since the quantum states of the cases (c) and (d) show different signs, they interfere destructively.}
\label{fig:4_BS}
\end{figure}

\subsection{Principle of the two-photon interference}\label{principle}
We consider two identical photons that impinge on a 50/50 beam splitter. The input state of the light field is given by $|\Psi_{in}\rangle = |1\rangle_{1H} |1\rangle_{2H}$, where the indices label the two input ports and the polarization of the photons. Here, we assume two photons of horizontal polarization. In the Schr\"{o}dinger picture we describe the evolution of the state using the unitary operator $\widehat{B}^\dagger$ as follows:
\begin{equation}
\widehat{B}^\dagger |1\rangle_{1H} |1\rangle_{2H} =
\widehat{B}^\dagger \hat{a}^\dagger_{1H} \hat{a}^\dagger_{2H} |0\rangle
\end{equation}
Since $\widehat{B}^{} \widehat{B}^\dagger$ is equal to the identity operator $\mathbf{1}$ and $\widehat{B}^\dagger |0\rangle = |0\rangle$ we can write
\begin{equation}
\widehat{B}^\dagger \hat{a}^\dagger_{1H} \hat{a}^\dagger_{2H} |0\rangle = 
\widehat{B}^\dagger \hat{a}^\dagger_{1H} \widehat{B}^{} ~ \widehat{B}^\dagger \hat{a}^\dagger_{2H} \widehat{B}^{} |0\rangle,
\end{equation}
which according to Eq.\,(\ref{eq:Entwicklung}) gives
\begin{equation}\label{eq:Schritt_1} 
\begin{array}{l}
\widehat{B}^\dagger \hat{a}^\dagger_{1H} \widehat{B}^{} ~ \widehat{B}^\dagger \hat{a}^\dagger_{2H} \widehat{B}^{} |0\rangle 
\\[2mm] 
= 
1/\sqrt{2} ~ (\hat{a}^\dagger_{1H} - \hat{a}^\dagger_{2H}) ~ 1/\sqrt{2} ~ (\hat{a}^\dagger_{1H} + \hat{a}^\dagger_{2H}) \, |0\rangle 
\\[2mm] 
= 
 1/2 ~ (\hat{a}^{\dagger 2}_{1H}  - \hat{a}^{\dagger 2}_{2H}
+ \hat{a}^{\dagger}_{1H} \hat{a}^{\dagger}_{2H} -
\hat{a}^{\dagger}_{2H} \hat{a}^{\dagger}_{1H}) \, |0\rangle.
\end{array}
\end{equation}
Each term in this sum of creation operators corresponds to one of four possible photon distributions in the beam splitter output ports, shown in Figure\,\ref{fig:4_BS}. In the first two cases, (a) and (b), both photons are found in either one or the other output, whereas in the cases (c) and (d), the photons go to different ports. The last two cases are indistinguishable, but the two expressions leading to cases (c) and (d) have opposite sign. Therefore the two possibilities interfere destructively. As a consequence, the two photons always leave the beam splitter as a pair and the output state is given by the superposition
\begin{equation}\label{eq:2_Photonen_Parallel}
\widehat{B}^\dagger |1\rangle_{1H} |1\rangle_{2H} =
\frac{1}{\sqrt{2}} (|2\rangle_{1H} |0\rangle_{2H} - |0\rangle_{1H}
|2\rangle_{2H}).
\end{equation}
This quantum interference occurs only if the photons are identical. If the photons were distinguishable, no interference takes place. For example, two photons of orthogonal polarization, $|\Psi_{in} \rangle = |1\rangle_{1H} |1\rangle_{2V}$, give rise to four different output states which are distinguishable by the photon polarization. In this case the overall output state can be written as a product state, e.g.,
\begin{equation}\nonumber
\begin{array}{l}
\widehat{B}^\dagger |1\rangle_{1H} |1\rangle_{2V} = \\[2mm]
\frac{1}{\sqrt{2}}( |1\rangle_{1H} |0\rangle_2 - |0\rangle_1
|1\rangle_{2H}) \otimes
\frac{1}{\sqrt{2}}(|1\rangle_{1V}|0\rangle_2 +
|0\rangle_1|1\rangle_{2V}),
\end{array}
\end{equation}
which describes the state of two independently distributed photons.

Note that all temporal aspects of the light field are neglected in the above discussion. In the next chapters, the two-photon interference is discussed under consideration of the photon duration and the time resolution of the detectors.

\subsection{Temporal aspects of the two-photon interference}\label{tempoarl_aspects}
We now take into account that the photodetections in the output ports of the beam splitter might occur at different times, $t_1$ and $t_2$. We use the Heisenberg picture to calculate the probability of a joint photodetection from the second-order correlation function,
\begin{equation}\label{eq:Korrelationsfunktion}
G^{(2)}(t_1, t_2) =  \sum_{s,s'} \mathrm{tr}[~\hat{\varrho}_{1,2}
~ \widehat{A}_{3s,4s'}(t_1,t_2)~],
\end{equation}
where $\hat{\varrho}_{1,2}$ describes the two-photon input state and the operator $\widehat{A}_{3s,4s'}(t_1,t_2)$ is given by
\begin{equation}
\begin{array}{l}
\widehat{A}^{}_{3s, 4s'}(t_1,t_2) := \hat{a}^\dagger_{3s}(t_1)
\hat{a}^\dagger_{4s'}(t_2) \hat{a}^{}_{4s'}(t_2)
\hat{a}^{}_{3s}(t_1) 
\\[2mm]
\mathrm{and} \quad s,s' \in \{H,V\}.
\end{array}
\end{equation}
The probability for a photodetection at the first detector within the time interval $[t_0 - \mathrm{d}t_0/2 , t_0 + \mathrm{d}t_0/2]$ and at the second detector within a time interval shifted by $\tau$, $[t_0 +\tau - \mathrm{d}\tau/2 , t_0 +\tau + \mathrm{d}\tau/2]$, is then given in analogy to Eq.\,(\ref{eq:detection_probability}),
\begin{equation}\label{eq:idealer_Photodetektor_2_neu}
P^{(2)}(t_0, \tau) = \eta_3 \eta_4 \!\!\! \int\limits_{t_0 -
\mathrm{d}t_0/2}^{t_0 + \mathrm{d}t_0/2} \!\! \mathrm{d}t_1 \!\!
\int\limits_{t_0 +\tau - \mathrm{d}\tau/2}^{t_0 +\tau +
\mathrm{d}\tau/2} \!\!\!\! \mathrm{d}t_2 ~~ G^{(2)}(t_1, t_2).
\end{equation}
Here we assume that the detectors have different efficiencies, $\eta_3$ and $\eta_4$. In analogy to Section \ref{sec:detect_probability} the smallest duration of the detection intervals is given by the detector time resolution, $T$, so that $\mathrm{d}t_0 \geq T$ and $\mathrm{d}\tau \geq T$. In the following, we calculate the joint detection probability in the limit of very short and very long photons.

If the photons are very short compared to the time resolution of the detectors, $\delta t << T$, the limits of the integration in Eq.\,(\ref{eq:idealer_Photodetektor_2_neu}) can be extended to infinity, so that
\begin{equation}\label{eq:no_timeresolution}
P^{(2)} =  \eta_3 \eta_4 \int \!\! \int \mathrm{d}t_1 ~
\mathrm{d}t_2 ~ G^{(2)}(t_1,t_2)
\end{equation}
gives the probability of a coincidence of photodetections.

For very long photons with $\delta t >> T$, the integration in Eq.\,(\ref{eq:idealer_Photodetektor_2_neu}) leads to
\begin{equation}
P^{(2)}(t_0,\tau) = \eta_3 \eta_4 ~ G^{(2)}(t_0, t_0 + \tau) ~
\mathrm{d}t_0 ~ \mathrm{d}\tau
\end{equation}
Therefore the probability of a joint photodetection can be studied as a function of the two detection times, $t_0$ and $t_0 + \tau$. In practice, only the time difference, $\tau$, between two photodetections is relevant. Therefore we integrate $P^{(2)}(t_0, \tau)$ over the time $t_0$ of the first photodetection. This gives
\begin{equation}\label{eq:P_Integration}
P^{(2)}(\tau) = \eta_3 \eta_4 ~T~ \int \mathrm{d}t_0 ~
G^{(2)}(t_0, t_0 + \tau),
\end{equation}
where $\mathrm{d}\tau$ is substituted by the detector time resolution $T$. The second-order correlation function, $G^{(2)}$, plays a central role in the calculation of the joint detection probability. It is now analyzed taking the polarization and the spatiotemporal modes of the photons into account.

\subsection{Correlation function}\label{correlation_function}
We calculate the correlation function $G^{(2)}$ for two photons characterized by two mode functions, $\xi_1$ and $\xi_2$. Without loss of generality, we assume that both photons are linearly polarized with an angle $\varphi$ between the two polarization directions. The state of the photons is then given by
$|1_{\xi_1}\rangle_{1H}$ and $\cos{\varphi}
~|1_{\xi_2}\rangle_{2H} |0\rangle_{2V} + \sin{\varphi}
~|0\rangle_{2H} |1_{\xi_2}\rangle_{2V}$, 
respectively. The density operator, $\hat{\varrho}^{}_{1,2} = |\Psi_{in} \rangle \langle\Psi_{in}|$, is given by the input state
\begin{equation}\label{eq:Eingangszustand}
|\Psi_{in}\rangle = \cos{\varphi} ~ |1_{\xi_1}\rangle_{1H}
|1_{\xi_2}\rangle_{2H} + \sin{\varphi} ~ |1_{\xi_1}\rangle_{1H}
|1_{\xi_2}\rangle_{2V},
\end{equation}
which is a superposition of the cases in which the impinging photons are parallel and perpendicular polarized to each other. The correlation function can then be written as a sum of two expressions $G^{(2)}_{HH}$ and $G^{(2)}_{HV}$,
\begin{equation}\label{eq:Korrelationsfunktion_phi}
G^{(2)} = \cos^2{\varphi} ~ G^{(2)}_{HH} + \sin^2{\varphi} ~
G^{(2)}_{HV},
\end{equation}
where the first function $G^{(2)}_{HH}$ accounts for the input state in which both photons have the same polarization and the second function $G^{(2)}_{HV}$ accounts for the perpendicular polarized case. Taking the mode functions into account, these two expressions read
\begin{eqnarray}
G^{(2)}_{HH}(t_1,t_2) &=& \frac{|\xi_1(t_1) \xi_2(t_2) -
\xi_2(t_1)
\xi_1(t_2)|^2}{4} \label{eq:Korrelationsfkt_Ergebnis_HH} \\
G^{(2)}_{HV}(t_1,t_2) &=& \frac{|\xi_1(t_1) \xi_2(t_2)|^2 +
|\xi_1(t_2) \xi_2(t_1)|^2}{4}.
\label{eq:Korrelationsfkt_Ergebnis_HV}
\end{eqnarray}
We emphasize that the correlation function for parallel polarized photons is always zero for $t_1 = t_2$, even if the mode functions $\xi_1(t)$ and $\xi_2(t)$ are not identical. As a consequence, the probability of a joint photodetection, Eq.\,(\ref{eq:P_Integration}), is always zero for $\tau = t_2 - t_1 =0$ , i.e. no simultaneous photodetections are expected even if the photons are distinguishable with respect to their mode functions.

As already mentioned in Chapter \ref{sec:frequency_modes}, the mode function can be written as the product of a real amplitude and a complex phase, $\xi_j(t) = \epsilon_j(t) \exp{(-i ~ \Phi_j(t))}$ with $j \in {1,2}$. Since the correlation function $G^{(2)}_{HV}$ for perpendicular polarized photons is independent of the phase, it can be written as
\begin{equation}\label{eq:Korrelationsfunktion_Ergebnis_HV_Ausgewertet}
G^{(2)}_{HV}(t_1,t_2) = \frac{(\epsilon_1(t_1) \epsilon_2(t_2))^2
+ (\epsilon_1(t_2) \epsilon_2(t_1))^2}{4}.
\end{equation}
This is not the case for the correlation function of parallel polarized photons which carries a phase-dependent interference term. It can be expressed as $G^{(2)}_{HH}(t_1,t_2) = G^{(2)}_{HV}(t_1,t_2) - F(t_1,t_2)$, with
\begin{widetext}
\begin{equation}\nonumber
F(t_1,t_2) := \frac{\epsilon_1(t_1) \epsilon_2(t_2)\epsilon_1(t_2)
\epsilon_2(t_1)}{2} \cos{(\Phi_1(t_1) - \Phi_1(t_2) + \Phi_2(t_2)
- \Phi_2(t_1) )}.
\end{equation}
\end{widetext}
However, this phase-dependency is only relevant, if $\Phi_1(t)$ and $\Phi_2(t)$ display a different time evolution. Otherwise the sum over the phases is always zero. Such a difference in the time evolution is given if, e.g., the frequencies of the photons are different. In that case, the interference term oscillates with the frequency difference, which gives rise to an oscillation in the joint photodetection probability, $P^{(2)}(\tau)$. This will be discussed further in Section \ref{sec:Zeit_4_Ornung}.

Taking the interference term into account, the overall correlation function, Eq.\,(\ref{eq:Korrelationsfunktion_phi}), can be summarized to
\begin{equation}\label{eq:Korrelationsfunktion_allgemein}
G^{(2)}(t_1,t_2) = G^{(2)}_{HV}(t_1,t_2) - \cos^2{\varphi} ~ F(t_1,t_2),
\end{equation}
where the effect of the interference term depends on the angle $\varphi$ between the two photon polarizations.

In the next two sections, the joint detection probability, Eq.\,(\ref{eq:idealer_Photodetektor_2_neu}), is analyzed for very long and very short photons.

\subsection{Two-photon interference without time resolution}\label{sec:HOM_Dip}
\begin{figure}[t]
\ \\
\begin{center}
\includegraphics[width=\columnwidth]{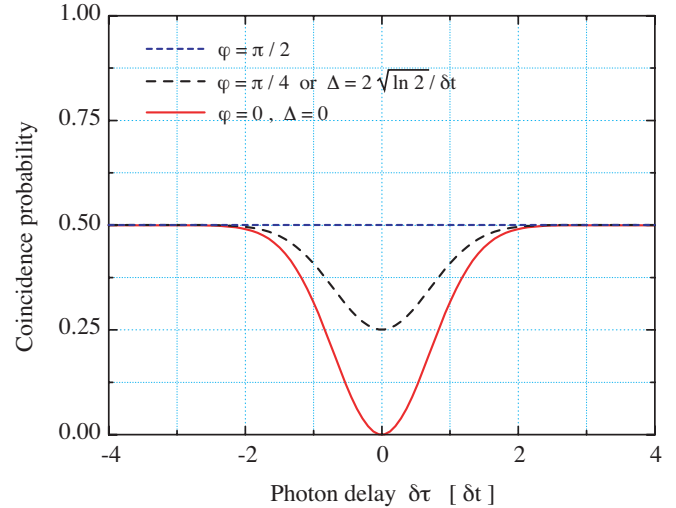}
\end{center}
\caption{Coincidence probability as a function of the arrival time delay, $\delta \tau$, of two linear polarized photons. In case of perpendicular polarized photons ($\varphi = \pi/2$) there is no interference at all and the coincidence probability shows the constant value $1/2$. If the photons are parallel polarized ($\varphi = 0$) and have identical frequency $(\Delta =0)$, there is a Gaussian-shaped dip which goes down to zero for simultaneous impinging photons, $\delta \tau =0$. Any difference in polarization or frequency leads to a reduced depth of this dip.}
\label{fig:Theorie-HOM-Dip}
\end{figure}

First we assume Gaussian-shaped photons which are very short compared to the time resolution of the photodetectors, $\delta t << T$. In this case one can only decide whether there is a coincidence of detections within the time interval $T$ or not, and the coincidence probability is given by Eq.\,(\ref{eq:no_timeresolution}). With a possible frequency difference $\Delta := \omega_{02} - \omega_{01}$ and an arrival-time delay $\delta \tau := \tau_{02} - \tau_{01}$ of the photons, the coincidence probability is given by
\begin{equation}\label{eq:HOM_dip}
P^{(2)} = \frac{1}{2} \left( 1 - \cos^2{\varphi} ~
\exp\left({-\frac{\delta t^2}{4/\Delta^2}}\right)
\exp\left({-\frac{\delta \tau^2}{\delta t^2}}\right)\right),
\end{equation}
where we assume that the photons hit perfect photodetectors with $\eta_3 = \eta_4 = 1$. We analyze the coincidence probability as a function of the photon delay $\delta\tau$ for different photon polarizations and frequency differences. This is shown in Figure\,\ref{fig:Theorie-HOM-Dip}.

As already discussed in Section \ref{principle}, perpendicular polarized photon pairs, $\varphi = \pi/2$, show no interference at all. Therefore, the probability for detecting photons at different output ports of the beam splitter is always $1/2$, independent of the photon delay, $\delta \tau$.

If the photons have identical polarizations, $\varphi =0$, and identical frequencies, $\Delta =0$, the coincidence probability shows a Gaussian-shaped dip centered at $\delta \tau = 0$. The minimum of $P^{(2)}$ is zero, indicating that the photons never leave the beam splitter through different output ports. If the photon pairs show any difference in their polarization, shape or frequency, there is no perfect interference and the minimum of the dip is no longer zero. Therefore the two-photon interference can be used to test the indistinguishability of photons.

The first measurement of the coincidence rate as a function of the relative photon delay was performed by Hong, Ou and Mandel \cite{Hong87} using photon pairs from a parametric downconversion source. They controlled the relative delay of the photons by shifting the position of the 50/50 beam splitter. The frequencies and bandwidths of the photons were adjusted by using two identical optical filters, so that the coincidence rate dropped nearly to zero for simultaneously impinging photons. As one can see from Eq.\,(\ref{eq:HOM_dip}), the width of the dip is identical to the photon duration, $\delta t$. Therefore this experiment was used to measure the duration and bandwidth of the photons.

So far, most two-photon interference experiments were performed with very short photons. Therefore the joint detection probability was only considered as a function of the photon delay, $\delta\tau$. However, if the photon duration is much larger than the detector time resolution, the time $\tau$ between the photodetections in the two output ports can be measured and the joint detection probability can additionally be analyzed in dependence of the detection-time difference.

\subsection{Time-resolved two-photon interference}\label{sec:Zeit_4_Ornung}
\begin{figure*}[t]
\begin{center}
\includegraphics[angle=0, width=\textwidth]{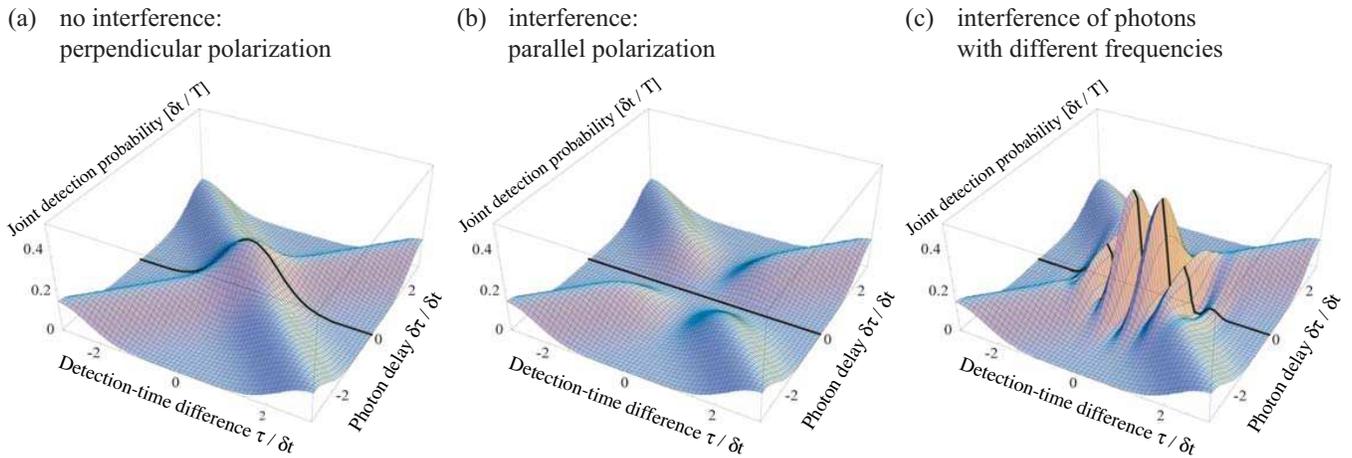}
\end{center}
\caption{Joint detection probability, $P^{(2)}$, as a function of the relative delay between the photons, $\delta\tau$, and the time difference between photodetections, $\tau$, for perpendicular polarized photons (a) and parallel polarized photons (b). In Figure\,(c) the parallel polarized photon pairs have a frequency difference $\Delta$, which leads to an oscillation in the joint detection probability. All times and frequencies are normalized by the photon duration, $\delta t$. }\label{fig:Theorie-3D}
\end{figure*}
We now assume, that the photon duration is much larger than the detection time resolution, $\delta t >> T$. Since the time, $\tau$, between the photodetections can be measured within the photon duration, the joint detection probability can be analyzed as a function of this detection-time difference. Using Eq.\,(\ref{eq:P_Integration}) and assuming Gaussian-shaped photons of identical duration, $\delta t$, the joint detection probability is given by
\begin{widetext}
\begin{equation}\label{eq:time_resolved}
P^{(2)}(\tau,\delta \tau) = \frac{T}{\sqrt{\pi} ~ \delta t} ~
\left[ \frac{1-\cos^2{\varphi}~\cos{(\Delta ~ \tau)}}{2} +
\sinh^2{\left(\frac{\tau ~ \delta\tau}{\delta t^2}\right)}
\right]  \cdot \exp\left({-\frac{\delta \tau^2 +
\tau^2}{\delta t^2}}\right).
\end{equation}
\end{widetext}
In Figure\,\ref{fig:Theorie-3D} the joint detection probability is shown as a function of the photon arrival-time delay, $\delta \tau$, and the detection time difference, $\tau$.

The sign of $\tau$ indicates which detector clicks first. Similarly, the sign of $\delta \tau$ determines which photon arrives first at the beam splitter. Note that the joint detection probability can only be different from zero if $|\tau| \approx |\delta \tau|$. This leads to the cross-like structure in Figure\,\ref{fig:Theorie-3D}\,(a-c). Since the photons only interfere if the relative delay is smaller than the photon duration, we focus our attention to the center of Figure\,\ref{fig:Theorie-3D}\,(a-c).

Again, we start our analysis with perpendicular polarized photon pairs. Obviously, no interference takes place, and as one can see in Figure\,\ref{fig:Theorie-3D}\,(a), even simultaneously impinging photons (with $\delta \tau = 0$) can be detected in different output ports of the beam splitter. The joint detection probability shows therefore a Gaussian-shaped peak. According to Eq.\,(\ref{eq:time_resolved}), the width of this peak is identical to the photon duration, $\delta t$. Since the photons are distinguishable by their polarization, an additional frequency difference, $\Delta$, does not affect this result. Assuming photon pairs with identical mode functions, the joint detection probability of perpendicular polarized photons can be used to determine the photon duration.

Figure\,\ref{fig:Theorie-3D}\,(b) shows the joint detection probability for parallel polarized photons of identical frequency, $\Delta =0$. For simultaneously impinging photons the joint detection probability is always zero, which indicates that the photons coalesce and leave the beam splitter always together.

If the parallel polarized photons show a frequency difference, the joint detection probability oscillates as a function of the detection time difference, $\tau$. This is shown in Figure\,\ref{fig:Theorie-3D}\,(c). As one can see from Eq.\,(\ref{eq:time_resolved}), the frequency difference, $\Delta$, determines the periodicity of this oscillation. We emphasize that the oscillation always leads to a minimum at $\tau =0$, independent of $\Delta$, so that even photons of different frequencies are never detected simultaneously in different output ports. Furthermore, the joint detection probability at the maxima is always larger than the joint detection probability for perpendicular polarized photons.

Without time resolution, the detection-time difference cannot be measured and the joint detection probability, $P^{(2)}(\tau,\delta\tau)$, has to be integrated over $\tau$. This links the results of a time-resolved two-photon interference to the discussion of Section \ref{sec:HOM_Dip}. In case of perpendicular polarized photons, the $\tau$-integrated function $P^{(2)}(\delta \tau)$ shows the constant value $1/2$. If the photons are identical, the integration leads to a Gaussian-shaped dip, which was already discussed in Section \ref{sec:HOM_Dip}. However, the oscillation in the joint detection probability for photon pairs with a frequency difference is no longer visible. The integration leads, in accordance to Eq.\,(\ref{eq:HOM_dip}), only to a reduced depth of the dip in $P^{(2)}(\delta\tau)$.

\section{Jitter}\label{sec:Jitter}
Up to now, we assumed that all photons of a given stream can be described by the same state vector $|1_{\xi}\rangle$. However, this requires a perfect single-photon source, which is able to generate a stream of photons without any variation in the parameters of the Gaussian mode functions. Here, we consider a more realistic scenario, in which a stream of single photons shows a jitter in the parameters, $\vartheta$. The quantum mechanical state of the photons is then given by the density operator of Eq.\,(\ref{eq:Dichtematrix}).

Such a jitter in the mode functions has important consequences on the results of measurements which can be performed on the single-photon stream. On the one hand, as already discussed in Section \ref{sec:detect_probability}, it affects the average detection probability of the photons in a way that its measurement does in general not reveal information about the duration or shape of each single photon. On the other hand, variations in the mode functions of photon pairs have an influence on the joint detection probability in two-photon interference experiments. This is discussed in some detail in the following two sections.

To analyze the effect of jitters on the two-photon interference, we consider two streams of Gaussian-shaped photons with a variation in the parameters of their mode functions. In analogy to Eq.\,(\ref{eq:Dichtematrix}) the density operator for photon pairs impinging on the beam splitter is given by
\begin{equation}\label{eq:two_photon_density_operator}
\hat{\varrho}_{1,2} = \int \!\!\! \int \mathrm{d}\vartheta_1 ~
\mathrm{d}\vartheta_2 ~ f_1(\vartheta_1) ~ f_2(\vartheta_2) ~
|1_{\xi_1}\rangle |1_{\xi_2}\rangle \langle1_{\xi_1}|
\langle1_{\xi_2}|,
\end{equation}
so that, using Eq.\,(\ref{eq:Korrelationsfunktion}), the correlation function reads
\begin{equation}\label{eq:correlationfunction_density_operator}
\begin{array}{l}
G^{(2)}(t_0, t_0+\tau) = 
\\[2mm]
\int \!\!\! \int \mathrm{d}\vartheta_{1}
\mathrm{d}\vartheta_{2} ~ f_1(\vartheta_{1}) f_2(\vartheta_{2}) ~~
\mathrm{tr}(\hat{\varrho}(\xi_1,\xi_2) ~ \hat{A}(t_0, t_0+\tau)).
\end{array}
\end{equation}
Here the expression $\hat{\varrho}(\xi_1,\xi_2)$ substitutes $|1_{\xi_1}\rangle |1_{\xi_2}\rangle \langle1_{\xi_1}| \langle1_{\xi_2}|$. In Eq.\,(\ref{eq:two_photon_density_operator}) we assumed that all photons have identical polarization and that they are completely independent from each other. Therefore the density operator has only diagonal elements. The parameters of the mode functions $\xi_1$ and $\xi_2$ of the two streams are summarized by $\vartheta_1$ and $\vartheta_2$, respectively. In general, all parameters of the mode functions could be subject to a variation, and all the variations could in principle depend on each other. However, in the following two sections, we focus our attention only on two examples of jitters and analyze the detection probability of photons, $P^{(1)}$, for a single photon stream as well as the joint detection probability, $P^{(2)}$, of two streams superimposed on a beam splitter.

First, in Section \ref{sec:frequency_jitter}, we consider streams of photons which are characterized by a variation of the center frequency, $\omega_{0j}$, so that each photon pair exhibits a variation of the frequency difference, $\Delta = \omega_{02} - \omega_{01}$. All remaining parameters of the mode functions, e.g. the duration of the photons, are assumed to be identical. In Section \ref{sec:emission_time_jitter}, we consider photons, which show only a variation in their emission time, so that photon pairs are characterized by a variation in their arrival-time delay $\delta \tau = \tau_{02} - \tau_{01}$.

\subsection{Frequency jitter}\label{sec:frequency_jitter}
We start our discussion of a frequency jitter by analyzing its effect on the average detection probability, $P^{(1)}(t_0)$, for a perfect photodetector with the detection efficiency $\eta =1$. If the frequency variation in the stream of single photons is described by a normalized distribution function, $f(\omega)$, the average detection probability is given, according to Eq.\,(\ref{eq:detection_probability_mixed}), by the integral
\begin{equation}
P^{(1)}(t_0) = T ~\int \mathrm{d}\omega ~ f(\omega) ~ |\xi(t_0,
\omega)|^2.
\end{equation}
Since only the phase of the Gaussian mode functions depends on the frequency, the absolute value, $|\xi(t_0,\omega)|^2 = \epsilon^2(t_0)$, is independent of $\omega$. Thus, the average detection probability is not affected by any frequency jitter and is entirely determined by the spatiotemporal mode function of each single photon.

\begin{figure}[t!]
\begin{center}
\includegraphics[angle=0, width=\columnwidth]{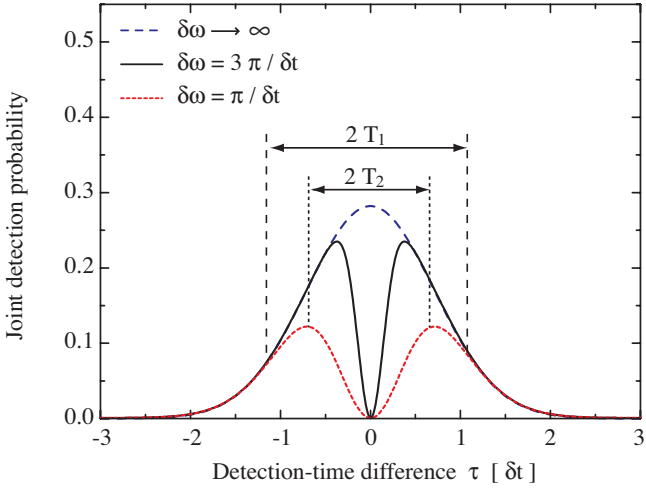}
\end{center}
\caption{Joint detection probability as a function of the detection-time difference, $\tau$, for simultaneously impinging photons, $\delta \tau =0$, of identical polarization. The photons are subject to a frequency jitter of width $\delta \omega$.}
\label{fig:Theorie-2D-inhom}
\end{figure}

However, a frequency jitter affects the joint detection probability of photon pairs superimposed on a beam splitter. To illustrate this, we assume two independent streams of photons, each fluctuating around a common center frequency $\omega_0$ according to a normalized Gaussian frequency distributions, $f_1(\omega_{01})$ and $f_2(\omega_{02})$. Hence, the frequency difference of the photon pairs, $\Delta = \omega_{02} - \omega_{01}$, shows also a normalized Gaussian variation,
\begin{equation}
f(\Delta) = \frac{1}{ \sqrt{\pi} ~\delta\omega} ~
\exp(-\Delta^2/\delta\omega^2),
\end{equation}
with width $\delta \omega$ depending on the widths of the frequency distributions of both streams, $\delta \omega = \sqrt{\delta \omega_{01}^2 + \delta \omega_{02}^2}$. The density operator of the photon pairs can then be expressed in terms of the distribution function of the frequency difference,
\begin{equation}
\hat{\varrho}_{1,2} = \int \mathrm{d}\Delta ~ f(\Delta) ~
\hat{\varrho}(\xi_1,\xi_2).
\end{equation}
As the operations of trace and integration are exchangeable, the correlation function, according to Eq.\,(\ref{eq:correlationfunction_density_operator}), can be written as
\begin{equation}\label{eq:Inhomogene_Verbreiterung}
G^{(2)}(t_0, \tau) = \int \mathrm{d}\Delta ~ f(\Delta) ~~
\mathrm{tr}(\hat{\varrho}(\xi_1,\xi_2) ~ \hat{A}(t_0,t_0+\tau)).
\end{equation}
For photons which are very long compared to the detector time-resolution, the joint detection probability is given by Eq.\,(\ref{eq:P_Integration}). In case of simultaneously impinging photons, $\delta \tau =0$, this leads to 
\begin{equation}\label{eq:frequency_jitter}
\begin{array}{l}
P^{(2)}(\tau) = 
\\[2mm]
\frac{T}{2 \sqrt{\pi} \delta t} ~ \left[ 1 -
\cos^2{\varphi}~
\exp{\left(-\frac{\tau^2}{4/\delta\omega^2}\right)} \right]
\exp\left({-\frac{\tau^2}{\delta t^2}}\right).
\end{array}
\end{equation}
For photons of parallel polarization, the result is shown in Figure\,\ref{fig:Theorie-2D-inhom}. In the limit of $\delta \omega \rightarrow \infty$ the joint detection probability shows a Gaussian-shaped peak of width $T_1 = \delta t$, which is the photon duration. As one can see from Eq.\,(\ref{eq:frequency_jitter}), the joint detection probability is always zero for $\tau=0$ as long as the width of the frequency distribution, $\delta \omega$, is finite. In fact, as one can deduce from Eq.\,(\ref{eq:frequency_jitter}), this leads to a dip in the joint detection probability around $\tau =0$ that is $T_2 = 2/\delta \omega$ wide. Note that $T_2$ represents a coherence time which must not be mixed up with the duration of each single photon, $\delta t$. 

To determine the amount of a frequency jitter from a time-resolved two-photon interference experiment, one has to perform two measurements. First, the joint detection probability of perpendicular polarized photons, $\varphi = \pi/2$, reveals the photon-duration, $\delta t$. Afterwards, the joint detection probability of parallel polarized photons, $\varphi = 0$, is used to measure $T_2$ and derive the width of the frequency jitter, $\delta \omega$. This is shown in detail in Chapter \ref{sec:Experiment}.

If the photons are very short compared to the detector time-resolution, the coincidence probability must be calculated using Eq.\,(\ref{eq:no_timeresolution}). The coincidence probability is then a function of the relative photon delay, $\delta \tau$, and is given by
\begin{equation}\label{eq:HOM_frequency_jitter}
P^{(2)}(\delta \tau) = \frac{1}{2} \left( 1 - \frac{2
\cos^2{\varphi}}{\sqrt{4 + \delta t^2 \delta \omega^2}} ~
\exp\left({-\frac{\delta \tau^2}{\delta t^2}}\right)\right).
\end{equation}
In analogy to Sec. \ref{sec:HOM_Dip}, a frequency jitter, $\delta \omega$, now leads to a decreased depth of the Gaussian-shaped dip, while the width of this dip is not affected and always identical to the photon duration.

In principle, it is possible to derive the frequency jitter also from a two-photon interference experiment without time-resolution, but there are some major disadvantages. First, the depth of the dip depends not only on a frequency jitter, but also on the mode matching of the transversal modes of both beams. A nonperfect mode matching leads to a factor comparable to $\cos^2{\varphi}$ in Eq.\,(\ref{eq:HOM_frequency_jitter}). Therefore, in contrast to the time-resolved measurement, one cannot distinguish between a nonperfect mode matching or a frequency jitter. Second, in case of two independent streams of photons from two different single-photon sources, it is impossible to decide whether a constant frequency difference or a frequency jitter is the reason for a decreased dip depth. And third, if the frequency jitter is large, the depth of the dip is very small, whereas in a time-resolved measurement, the dip-depth remains unchanged. As it is much more reliable to determine a small width rather than a small depth, the time-resolved method is much more powerful.

\subsection{Emission-time jitter}\label{sec:emission_time_jitter}
Now we assume a stream of single photons which shows a jitter in the emission time of each photon. This variation of the emission time is assumed to be given by a normalized Gaussian distribution function, $f(\tau_0)$. The average detection probability of the photons for an ideal photodetector with $\eta=1$ is again given by Eq.\,(\ref{eq:detection_probability_mixed}),
\begin{equation}\label{eq:detection_probability_time_jitter}
P^{(1)}(t_0) = T ~  \int \!\! \mathrm{d}\tau ~ f(\tau) ~ |\xi(\tau
- t_0)|^2.
\end{equation}
This is a convolution of the detection probability, $|\xi(t_0)|^2$, of each single photon and the emission-time distribution, $f(\tau_0)$, of the photon stream. Therefore the average detection probability is always broader than the detection probability which would arise solely from the duration of each single photon. This shows that a variation in the parameters of the spatiotemporal mode functions can alter the detection probability of the photons. Therefore, in general, the average detection probability is not identical to the detection probability of individual photons.

\begin{figure}[t!]
\begin{center}
\includegraphics[angle=0, width=\columnwidth]{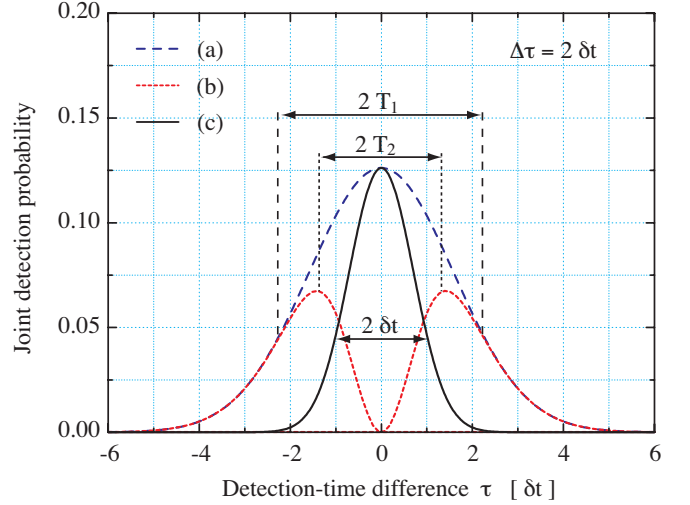}
\end{center}
\caption{Joint detection probability as a function of the detection time difference, $\tau$, for photon pairs with a variation in their relative photon delay, $\delta \tau$. The photons are assumed to be (a) perpendicular and (b) parallel polarized to each other. Curve (c) is a Gaussian mode function of width $\delta t$. The width, $T_1$, of the Gaussian peak in (a) is broadened by the variation of the relative photon delay, $\Delta \tau$, which is here assumed to be $2 ~\delta t$.} \label{fig:Theorie_Anfangszeit_2D}
\end{figure}

To investigate the influence of an emission-time jitter on the joint detection probability in a two-photon interference experiment, we now assume two streams of photons with a Gaussian emission-time distribution of identical width, $\Delta\tau$. In this case, the photon pairs are characterized by a jitter in the arrival-time delay of the photons, which is again given by a Gaussian distribution,
\begin{equation}
f(\delta\tau) = \frac{1}{ \sqrt{\pi} ~\Delta\tau} ~ \exp(-
\delta\tau^2/\Delta\tau^2).
\end{equation}
The correlation function $G^{(2)}(t_0, t_0 + \tau)$ can be written in analogy to Eq.\,(\ref{eq:Inhomogene_Verbreiterung}), using only the variation of the relative photon delay,
\begin{equation}
G^{(2)}(t_0, \tau) = \int \mathrm{d}(\delta\tau) ~ f(\delta\tau)
~~ \mathrm{tr}(\hat{\varrho}(\xi_1,\xi_2) ~
\hat{A}(t_0,t_0+\tau)),
\end{equation}
and the joint detection probability has to be calculated according to Eq.\,(\ref{eq:P_Integration}). In case of simultaneously impinging photons, this leads to
\begin{widetext}
\begin{equation}\label{eq:emission-time_jitter}
P^{(2)}(\tau)= \frac{T}{2 \sqrt{\pi} \sqrt{\delta t^2 + \Delta
\tau^2}} ~ \left[ 1 - \cos^2{\varphi}~
\exp\left({-\frac{\tau^2}{\delta t^2 + \delta t^4 /
\Delta\tau^2}}\right) \right]  \cdot \exp\left({-\frac{\tau^2}{\delta t^2 +
\Delta\tau^2}}\right).
\end{equation}
\end{widetext}
In contrast to the previous case, now the width of the Gaussian-shaped peak in the joint detection probability of perpendicular polarized photons is no more identical to the photon duration. The variation in the emission time affects also the amplitude of the spatiotemporal mode functions and affects the joint detection probability even without interference. This can be seen in Figure\,\ref{fig:Theorie_Anfangszeit_2D}, which shows the joint detection probability for photon pairs characterized by a distribution of the photon arrival-time delay. The width, $T_1$, of the Gaussian-shaped peak is no longer identical to the photon duration, $\delta t$. As one can derive from Eq.\,(\ref{eq:emission-time_jitter}), it is now broadened by the width of the variation in the photon delay, $\Delta \tau$ :
\begin{equation}
T_1 = \sqrt{\delta t^2 + \Delta \tau^2}.
\end{equation}
Furthermore, the width of the dip in case of parallel polarized photons is not independent of the width $T_1$, but is given by
\begin{equation}
T_2 = \sqrt{\delta t^2 + \delta t^4 / \Delta\tau^2} = \frac{\delta
t}{\Delta \tau} ~T_1.
\end{equation}
A time-resolved two-photon interference experiment can again be used to determine the variation in the emission time of the photon streams. However, since the shapes of the joint detection probabilities for a frequency and an emission-time jitter are identical, it is in general not possible to distinguish between the two. Nonetheless, one can determine the maximum values of both jitters, as well as all pairs of frequency and an emission-time jitters matching the data. This is discussed in the next chapter.

Note that in case of very short photons, the coincidence probability has again to be calculated using Eq.\,(\ref{eq:no_timeresolution}). The width of the Gaussian-shaped dip in the coincidence probability
\begin{equation}
P^{(2)}(\delta \tau) = \frac{1}{2} \left( 1 -
\frac{\cos^2{\varphi}}{\sqrt{1 + \Delta \tau^2 / \delta t^2}} ~
\exp\left({-\frac{\delta \tau^2}{\delta t^2 + \Delta \tau^2
}}\right)\right).
\end{equation}
is broadened by the emission-time jitter, $\Delta \tau$, and the depth of the dip is also decreased by $\Delta \tau$. Again it is possible to determine the emission-time jitter from such a two-photon interference experiment without time resolution. The disadvantages of such a procedure have already been discussed in Section \ref{sec:frequency_jitter}.

\subsection{Autocorrelation function of the photon's shape}\label{sec:autocorrelation_function}
We now consider only one source of single photons that we want to characterize using a two-photon interference experiment. We assume that the stream of photons generated by this source is split up in such a way that each single photon is randomly directed along two different paths. These two paths are of different length and the repetition rate of the source is chosen in such a way that only successively generated photons impinge on the beam splitter at the same time. The details of such an experiment are discussed in Chapter \ref{sec:Experiment}.

In general, we must distinguish between the photon ensemble of the whole stream and the subensemble of photon pairs superimposed on the beam splitter. The latter consists only of successively generated photons and the characterization of a single-photon source by a two-photon interference experiment takes into account only this subensemble. Since the jitter in the subensemble of successive photon pairs does not have to be identical to the jitter in the whole photon stream, we need a method to decide whether the results of a two-photon interference experiment can be generalized to all photons generated by the single-photon source. In case of very long photons, this can be done by comparing the joint detection probability, $P^{(2)}(\tau)$, for perpendicular polarized photons and the autocorrelation function, $A^{(2)}(\tau)$, of the average detection probability of the whole photon stream.

We start our discussion with a stream of identical single photons, so that the average detection probability is simply given by the square of the amplitude of the spatiotemporal mode function, $P^{(1)}(t_0) = T~ \epsilon^2(t_0)$. On the one hand, the autocorrelation function of $P^{(1)}(t_0)$ reads
\begin{equation}\label{eq:autocorrelation_function}
\begin{array}{l}
A^{(2)}(\tau) = 
\\[2mm]
\int \mathrm{d}t_0 ~ P^{(1)}(t_0)
P^{(1)}(t_0+\tau) =  T^2 \int \mathrm{d}t_0 ~ (\epsilon(t_0)
\epsilon(t_0+\tau))^2.
\end{array}
\end{equation}
On the other hand, the joint detection probability for perpendicular polarized photon pairs of this stream is given by Eq.\,(\ref{eq:Korrelationsfunktion_Ergebnis_HV_Ausgewertet}), which leads to
\begin{equation}
P^{(2)}(\tau) \propto T^2 \int \mathrm{d}t_0 ~ (\epsilon(t_0)
\epsilon(t_0+\tau))^2.
\end{equation}
Therefore the joint detection probability for perpendicular polarized photons and the autocorrelation function have the same shape.

However, if the photons show a variation in their spatiotemporal modes, the two functions are no longer equal. In the following, we assume that the variations in the whole photon stream are described by a normalized distribution function $f(\vartheta)$, whereas the variations in the subensemble of successively emitted photons is given by $\tilde{f}(\vartheta)$. In general, both functions do not have to be identical, i.e. the jitter in the subensemble can be smaller than the jitter in the whole photon stream. The autocorrelation function of the average detection probability $P^{(1)}(t_0)$, see Eq.\,(\ref{eq:detection_probability_mixed}), is given by
\begin{equation}\label{eq:Autokorrelation}
\begin{array}{l}
A^{(2)}(\tau) = 
\\[2mm]
T^2 \int \!\! \mathrm{d} t_0  ~ \int \!\!\! \int
\mathrm{d}\vartheta_1 \mathrm{d}\vartheta_2 ~ f(\vartheta_1)
f(\vartheta_2) ~(\epsilon(t_0,\vartheta_1)
\epsilon(t_0+\tau,\vartheta_2))^2.
\end{array}
\end{equation}
The joint detection probability,
\begin{equation}\label{eq:Kreuzkorrelation}
\begin{array}{l}
P^{(2)}(\tau) \propto 
\\[2mm]
T^2 ~ \int \!\! \mathrm{d} t_0 \int \!\!\!
\int \mathrm{d}\vartheta_1 \mathrm{d}\vartheta_2 ~
\tilde{f}(\vartheta_1) \tilde{f}(\vartheta_2) ~
(\epsilon(t_0,\vartheta_1) \epsilon(t_0+\tau,\vartheta_2))^2
\end{array}
\end{equation}
is therefore only equal to $A^{(2)}(\tau)$, if $(\epsilon(t_0,\vartheta_1) \epsilon(t_0+\tau,\vartheta_2))^2$ is either independent of $\vartheta_{1}$ and $\vartheta_{2}$, or if the distribution function $f(\vartheta)$ of the whole photon stream is identical to the distribution function $\tilde{f}(\vartheta)$ of successively emitted photons.

The comparison of the joint detection probability of perpendicular polarized photon pairs and the autocorrelation function of the average detection probability therefore answers the question whether the results of a two-photon interference experiment can be generalized to the whole photon stream.

\section{Experiment and Results}\label{sec:Experiment}
In the previous three chapters we discussed the theoretical background for characterizing single photons using two-photon interference. Now, we show how to use this method to experimentally characterize single photons that are emitted from only one source. This single-photon source has been realized using vacuum-stimulated Raman transitions in a single Rb atom located inside a high-finesse optical cavity. In Section \ref{sec:setup} we briefly review the principle of this source and discuss the experimental setup, which was used to investigate the two-photon interference. For further details concerning the single-photon generation, we refer to Kuhn et al. \cite{Kuhn02:2} and references therein. The measurement of the average detection probability of a stream of photons emitted from this source is discussed in Section \ref{sec:Pulsform}. As already discussed, we use the autocorrelation function of the average detection probability to determine whether the results of a two-photon interference experiment can be generalized to the whole photon stream. Since the duration of the photons is much longer than the time resolution of the detectors, the interference of successively emitted photons is measured in a time-resolved manner. The results and the interpretation of these measurements are discussed in detail in Section \ref{sec:Ergebnis-Interferenz} and
\ref{sec:interpretation}.

\subsection{Single-photon source and experimental setup}\label{sec:setup}
\begin{figure}[!t]
\begin{center}
\includegraphics[angle=0, width=1\columnwidth]{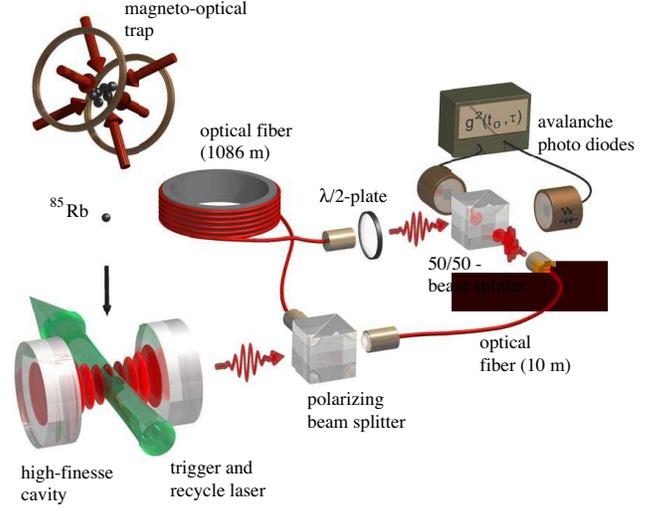}
\end{center}
\caption{Single-photon source and experimental setup used to investigate the two-photon interference of successively emitted photons. The photons are generated in an atom-cavity system by an adiabatically driven stimulated Raman transition. A polarizing beam splitter directs the photons randomly into two optical fibers. The delay from photon to photon matches the travel-time difference in the two fibers, so that successively emitted photons can impinge simultaneously on the 50/50 beam splitter. Using a half-wave plate, the polarization of the photons can be chosen parallel or perpendicular to each other. The photons are detected using avalanche photodiodes with a detection efficiency of 50\,\% and a dark-count rate of 150\,Hz.} \label{fig:Aufbau-Experiment-3D}
\end{figure}
A sketch of the single-photon source and the experimental setup that we use to superimpose successively generated photons on a 50/50 beam splitter is shown in Figure \ref{fig:Aufbau-Experiment-3D}. The single-photon generation starts with $^{85}$Rb atoms released from a magneto-optical trap. The atoms enter the cavity mostly one-at-a-time (the probability of having more than one atom is negligible). Each atom is initially prepared in $|e\rangle \equiv |5S_{1/2}, F=3\rangle$, while the cavity is resonant with the transition between $|g\rangle\equiv |5S_{1/2}, F=2\rangle$ and $|x\rangle\equiv |5P_{3/2}, F=3\rangle$. On its way through the cavity, the atom experiences a sequence of laser pulses that alternate between triggering single-photon emissions and repumping the atom to state $|e\rangle$: The $2\,\mu$s-long trigger pulses are resonant with the $|e\rangle\leftrightarrow |x\rangle$ transition and drive an adiabatic passage (STIRAP) to $|g\rangle$ by linearly increasing the Rabi frequency. This transition goes hand-in-hand with a photon emission. In the ideal case, the duration and pulse shape of each photon depend in a characteristic manner on  the temporal shape and intensity of the triggering laser pulses \citep{Keller04}. As we will discuss in Section \ref{sec:Ergebnis-Interferenz}, the photon-frequency can be chosen by an appropriate frequency of the trigger laser \citep{Legero04}. Between two photon emissions, another laser pumps the atom from $|g\rangle$ to $|x\rangle$, from where it decays back to $|e\rangle$. While a single atom interacts with the cavity, the source generates a stream of single photons one-after-the-other. The efficiency of the photon generation is 25\,\%. 

As described in detail by Legero \cite{Legero05}, the source has been optimized with respect to jitters by compensating the Earth's magnetic field inside the cavity and by adding to the recycling scheme an additional $\pi$-polarized laser driving the transition $|5S_{1/2}, F=3\rangle \leftrightarrow |5P_{3/2}, F=2\rangle$ to produce a high degree of spin-polarization in $5S_{1/2}, F=3, m_{F}=\pm 3$. This results in an increased coupling of the atom to the cavity. We have characterized the emitted photons by two-photon interference measurements before and after this optimization.

To superimpose two successively emitted photons on the 50/50 beam splitter, they are directed along two optical paths of different length. These paths are realized using two polarization maintaining optical fibers with a length of 10 m and 1086 m, respectively. Since the photon polarization is \`{a} priori undefined, a polarizing beam splitter is used to direct the photons randomly into the long or short fiber. The time between two trigger pulses is adjusted to match the travel-time difference of the photons in the two fibers, which is $\Delta t = 5.28\,\mu$s. With a probability of 25\,\%, two successively emitted photons therefore impinge on the beam splitter simultaneously. In addition, we use a half-wave plate to adjust the mutual polarization of the two paths.

\subsection{Average detection probability}\label{sec:Pulsform}

\begin{figure}[t!]
\begin{center}
\includegraphics[angle=0, width=\columnwidth]{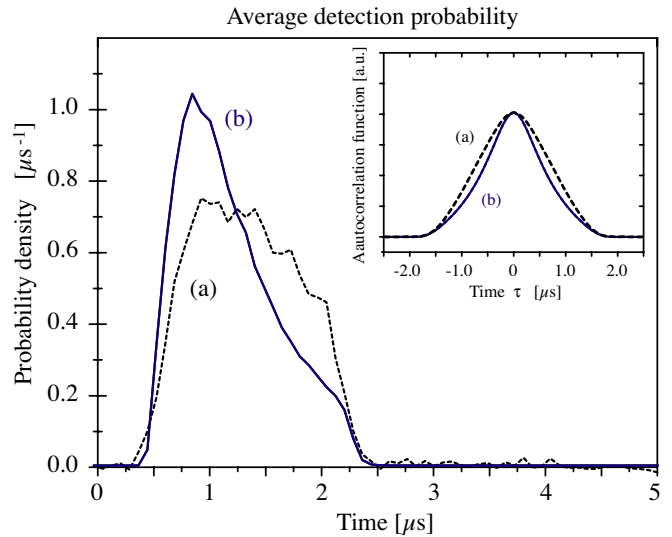}
\end{center}
\caption{Probability density for photodetections averaged over $10^3$ photons (a) before and (b) after optimizing the single-photon source. The data are corrected for the detector dark-count rate. The inset shows the corresponding autocorrelation functions, $A^{(2)}(\tau)$.}\label{fig:Photon_arrivals_2}
\end{figure}

First, we investigate the average detection probability of the photon stream. For this measurement, the long fiber is closed and the detection times of about $10^3$ photons are recorded with respect to their trigger pulses. From these photons, we calculate the probability density for a photodetection, shown in Figure \ref{fig:Photon_arrivals_2}. The measurement has been done (a) before and (b) after optimizing the single-photon source. The autocorrelation functions of both curves, shown in the inset of Figure \ref{fig:Photon_arrivals_2}, were calculated using Eq.\,(\ref{eq:autocorrelation_function}). The width of these curves is (a) $1.07\,\mu$s and (b) $0.81\,\mu$s. In Section \ref{sec:interpretation} we compare both results with the joint detection probability of perpendicular polarized photon pairs.

Note that one obtains no information on the shape or duration of individual photons from a detection probability that is averaged over many photodetections. As already stated in Section \ref{sec:emission_time_jitter}, such a measurement does not exclude that the photons are very short so that the average probability distribution reflects only an emission-time jitter. Only from a time-resolved two-photon interference experiment, one obtains information on the duration of the photons. This is discussed in the next section.

\subsection{Time-resolved two-photon interference}\label{sec:Ergebnis-Interferenz}
The detection times of about $10^5$ photons are registered by the two detectors in the output ports of the beam splitter,  while the photons in each pair impinge simultaneously, i.e. with $\delta \tau =0$. The number of joint photodetections, $N^{(2)}$, is then determined from the recorded detection times as a function of the detection-time difference, $\tau$ (using 48\,ns to 120\,ns long time bins). To do that, the photon duration must exceed the time resolution of the detectors. Otherwise,  joint detection probabilities could only be examined as a function of the arrival-time delay, $\delta \tau$, like in most other experiments.

\begin{figure}[!t]
\begin{center}
\includegraphics[angle=0, width=0.97\columnwidth]{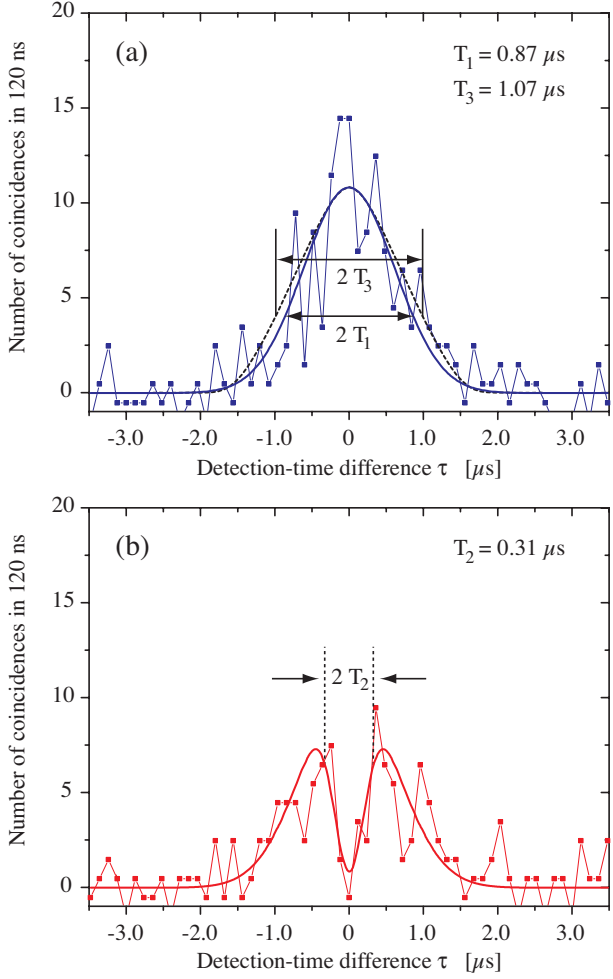}
\end{center}
\caption{Number of joint photodetections in 120\,ns-long time bins as a function of the detection-time difference, $\tau$, before optimizing the single-photon source. The results are shown for photons of (a) perpendicular polarization and (b) parallel polarization. In both cases, the data is accumulated for a total number of 73000 photodetections. The solid lines are numerical fits of the theoretical expectations to the data. $T_1$ is the width of the Gaussian peak in (a), and the width of the dip for parallel polarized photons (b) is given by $T_2$. The dotted curve shows the $T_3$-wide autocorrelation function, $A^{(2)}(\tau)$, of the average detection probability.} 
\label{fig:Messung-HOM-alt-Kontrast}
\end{figure}

\begin{figure}[!t]
\begin{center}
\includegraphics[angle=0, width=0.97\columnwidth]{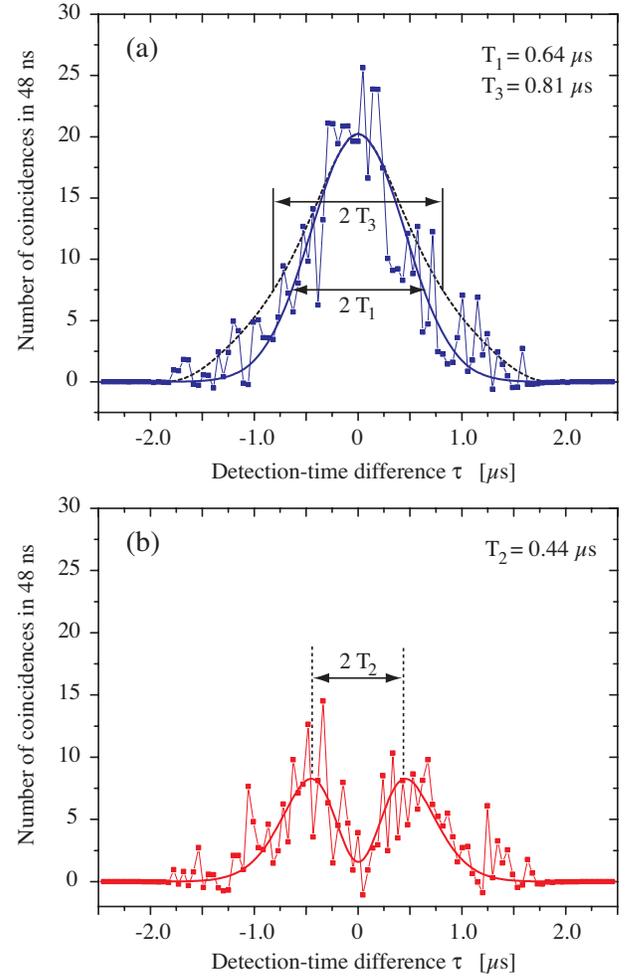}
\end{center}
\caption{Number of joint photodetections in 48 ns long time bins after optimizing the single-photon source. The data is shown in (a) for perpendicular and in (b) for parallel polarized photons. In both cases, the data is accumulated for a total number of 139000 photodetections. The dotted curve shows the $T_3$-wide autocorrelation function of the average detection probability. Compared to the results of Figure \ref{fig:Messung-HOM-alt-Kontrast}, the width of the Gaussian peak in (a) is decreased to $T_1=0.64\,\mu$s and it is clearly smaller than $T_3=0.81\,\mu$s. With parallel polarizations, a dip of increased width,  $T_2=0.44\,\mu$s, is observed. All results are discussed in Section \ref{sec:interpretation}.}
\label{fig:Messung-HOM-neu-Kontrast}
\end{figure}

We have performed these two-photon interference  experiments before and after optimizing the single-photon source. The results are shown in Figure \ref{fig:Messung-HOM-alt-Kontrast} and \ref{fig:Messung-HOM-neu-Kontrast}, respectively. Each experiment is first performed with photons of (a) perpendicular and then with photons of (b) parallel polarization, until about $10^5$ photons are detected. Note that the number of joint photodetections is corrected for the constant background contribution stemming from detector dark counts.

In case of perpendicular polarization, no interference takes place. In accordance with Section \ref{sec:Zeit_4_Ornung}, the number of joint photodetections shows a Gaussian peak centered at $\tau=0$. This signal is used as a reference, since any interference leads to a significant deviation. If we now switch to parallel polarization, identical photons are expected to leave the beam splitter as a pair, so that their joint detection probability should be zero for all values of $\tau$. In the experiment, however, the signal does not vanish completely. Instead, we observe  a pronounced minimum around $\tau =0$, which complies well with the behavior one expects for varying spatio-temporal modes, as shown in Figures \ref{fig:Theorie-2D-inhom} and \ref{fig:Theorie_Anfangszeit_2D}.  In analogy to Eq.\,(\ref{eq:frequency_jitter}) and (\ref{eq:emission-time_jitter}), respectively, the number of joint photodetections is then given by
\begin{equation}\label{eq:number_joint_detections}
N^{(2)}(\tau) =  N^{(2)}_0 ~
\exp\left({-\frac{\tau^2}{T^2_1}}\right) ~\left[ 1 -
\cos^2{\varphi}~ \exp{\left(-\frac{\tau^2}{T^2_2}\right)} \right].
\end{equation}
$N^{(2)}_0$ is the peak value at $\tau=0$ that we measure for perpendicular polarized photons, $\varphi=\pi/2$. The time $T_1$ is the width of this Gaussian peak, whereas $T_2$ gives the width of the dip for photons of parallel polarization, $\varphi=0$. In Section \ref{sec:interpretation}, both numbers are used to deduce the frequency and the emission-time jitter. The two widths, $T_1$ and $T_2$, are obtained from a fit of Eq.\,(\ref{eq:number_joint_detections}) to the measured data. This is done in two steps.  First, we obtain $T_1$ and $N^{(2)}_0$ from a fit to the data taken with perpendicular polarized photons. We then keep these two values and obtain the dip width $T_2$ from a subsequent fit to the data with parallel polarized photons. In this second step, we use a polarization term of $\cos^2{\varphi}=0.92$ to take into account that we have a small geometric mode mismatch. This is well justified since mode mismatch and non-parallel polarizations affect the signal in the same manner. The value of $\varphi$ has been obtained from an independent second-order interference measurement.

\begin{figure}[!t]
\begin{center}
\includegraphics[angle=0, width=\columnwidth]{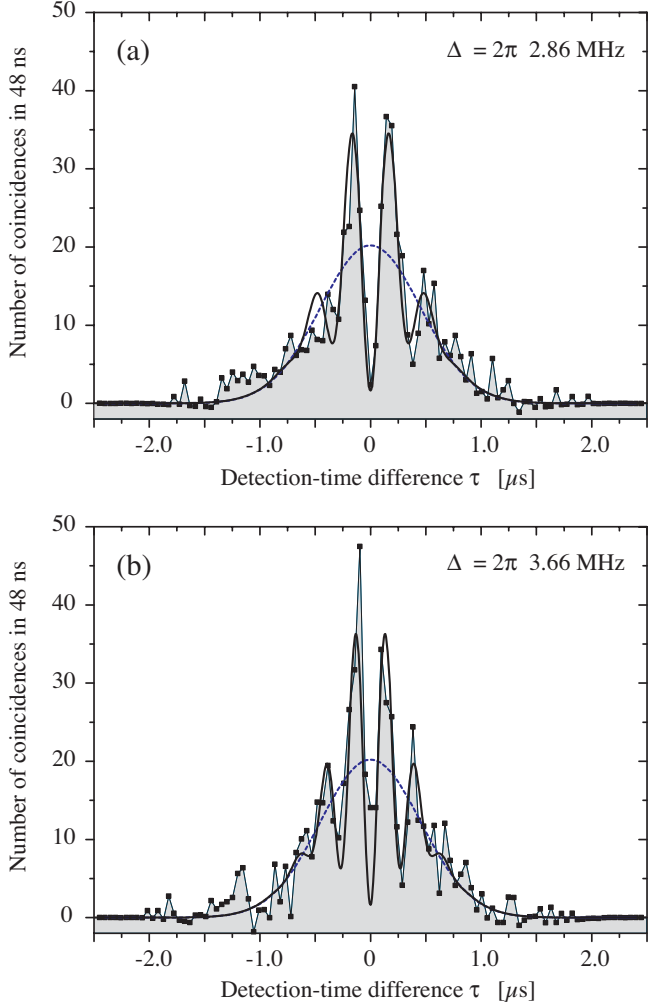}
\end{center}
\caption{Two-photon interference of parallel polarized photons with a frequency difference, $\Delta$, of (a) $2 \pi \times 2.8\,$MHz and (b) $2 \pi \times 3.8\,$MHz. The number of joint photodetections is accumulated over (a) 210000 and (b) 319000 detection events. It oscillates as a function of the detection-time difference. The solid curves represent numerical fits to the data with a frequency-difference of (a) $\Delta = 2\pi \times 2.86\,$MHz and (b) $\Delta = 2 \pi \times 3.66\,$MHz. The dotted curve shows the reference signal measured with perpendicular polarized photons.} \label{fig:Messung-HOM-3-4-MHz}
\end{figure}

As one can see by comparing Figures \ref{fig:Messung-HOM-alt-Kontrast} and \ref{fig:Messung-HOM-neu-Kontrast}, the compensation of the Earth's magnetic field and the improved recycling scheme lead to a decreased width $T_1$ of the Gaussian peak in (a) and a broader dip $T_2$ in (b). As we discuss in the following, these results show that this optimization has successfully reduced the jitter in the mode function of the photons.

Moreover, as shown in Figure \ref{fig:Messung-HOM-3-4-MHz}(a) and (b), we resolve a pronounced oscillation in the number of joint photodetections when a frequency difference, $\Delta$, is deliberately introduced between the interfering photons \citep{Legero04}. This is achieved by driving the atom-cavity system with a sequence of trigger pulses that alternate between two frequencies. The frequency difference between consecutive pulses is either (a) $2 \pi \times 2.8\,$MHz or (b) $2 \pi \times 3.8\,$MHz. In accordance with Section \ref {sec:Zeit_4_Ornung}, the oscillation in the joint detection probability always starts with a minimum at $\tau=0$, and the maxima exceed the reference signal that we measure with perpendicular polarized photons.

The latter experiment has been performed with the optimized single-photon source. The number of detected photons (corrected for the number of dark counts) equals the photon number in Figure \ref{fig:Messung-HOM-neu-Kontrast}. Therefore $N^{(2)}_0, T_1$ and $T_2$ are well known from this previous measurement. As shown by Legero \cite{Legero05}, the only remaining parameter one can obtain from a fit of the joint detection probability,
\begin{equation}\label{eq:number_joint_detections_oscillate}
\begin{array}{l}
N^{(2)}(\tau) =  
\\[2mm]
N^{(2)}_0 ~
\exp\left({-\frac{\tau^2}{T^2_1}}\right) ~\left[ 1 -
\cos^2{\varphi}~ \cos{(\tau ~ \Delta)} ~
\exp{\left(-\frac{\tau^2}{T^2_2}\right)} \right],
\end{array}
\end{equation}
to the data is the frequency difference  $\Delta$. Its fit value agrees very well  with the frequency differences that we imposed on consecutive pulses. We therefore conclude that  the adiabatic Raman transition we use to generate the photons  allows us to adjust the  single-photon frequency. Moreover, the oscillations in the joint detection probability impressively demonstrate that time-resolved two-photon interference experiments are able to reveal small phase variations between the photons, like, e.g., the frequency difference we have deliberately imposed  here.

\subsection{Interpretation of the results}\label{sec:interpretation}
We start our analysis by comparing the autocorrelation function of the average detection probability with the result of the two-photon coincidence measurement with perpendicular polarized photons, shown in Figures \ref{fig:Messung-HOM-alt-Kontrast}(a) and \ref{fig:Messung-HOM-neu-Kontrast}(a). The shape of the autocorrelation function, $A^{(2)}(\tau)$, is commensurable with the data obtained in the two-photon correlation experiment before the optimization of the source, but it differs significantly afterwards.  As discussed in Section \ref{sec:autocorrelation_function}, the different widths of both curves, $T_1 = 0.64\,\mu$s and $T_3 = 0.81\,\mu$s, indicate that the photon stream is subject to variations in the spatiotemporal mode functions that are much less pronounced in the subensemble of consecutive photons. These variations cannot be attributed to a jitter in the photon frequency, since the autocorrelation function and the joint detection probability, given by Eq.\,(\ref{eq:Autokorrelation}) and (\ref{eq:Kreuzkorrelation}), depend only on the frequency-independent amplitude of the mode functions. Therefore the emission time and/or the duration of the photons must be subject to a jitter. As a consequence, the average detection probability shown in Figure \ref{fig:Photon_arrivals_2} cannot represent the shape of the underlying single-photon wavepackets. In particular the width of the measured photon detection probability is broadened due to the emission-time jitter.

Moreover, the discrepancy between $A^{(2)}(\tau)$ (Figure \ref{fig:Photon_arrivals_2}) and the Gaussian peak (Figures \ref{fig:Messung-HOM-alt-Kontrast} (a) and \ref{fig:Messung-HOM-neu-Kontrast} (a)) shows that the variations in the whole photon stream are larger than the variations in the subensemble of consecutive photons. Therefore the following analysis of the times $T_1$ and $T_2$ of the two-photon interference cannot be generalized to the whole photon stream.

\begin{figure}[!t]
\begin{center}
\includegraphics[angle=0, width=\columnwidth]{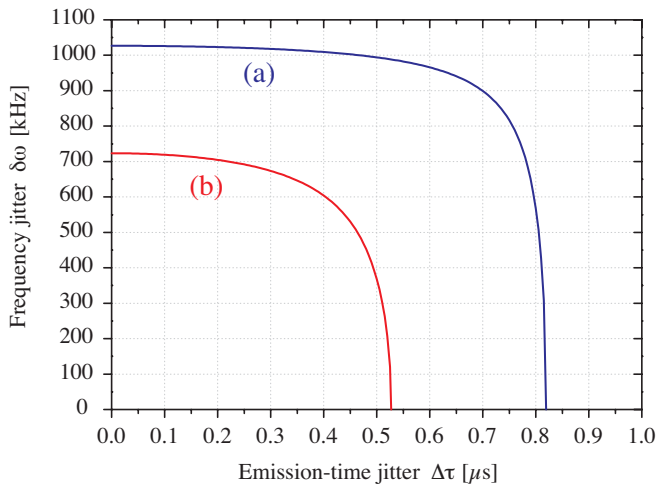}
\end{center}
\caption{Frequency jitter, $\delta \omega$, and emission-time jitter, $\Delta \tau$, before (a) and after (b) optimizing the single-photon source. The two curves represent all pairs of jitters that match the two widths,  $T_1$ and $T_2$, found in the two-photon interference experiments.} \label{fig:Parameter-Raum}
\end{figure}

To figure out the photon characteristics that can explain the measured quantum-beat signal, we restrict the analysis of $T_1$ and $T_2$ to a frequency and an emission-time jitter. If we assume that successively emitted photons show only a variation of their frequencies, then $T_1$ is identical to the photon duration, whereas $T_2 = 2/\delta\omega$ is solely due to the frequency variation. In this case, $T_2$ is identical to the coherence time, which one could also measure using second-order interference \citep{Santori02,Jelezko03}. As shown in Figures \ref{fig:Messung-HOM-alt-Kontrast} and \ref{fig:Messung-HOM-neu-Kontrast}, optimizing the single-photon source reduces the bandwidth of the frequency variation from $\delta \omega / 2 \pi =1.03\,$MHz to $\delta \omega / 2 \pi =720$\,kHz. The remaining inhomogeneous broadening of the photon frequency can be attributed to several technical reasons. First, static and fluctuating magnetic fields affect the energies of the Zeeman sublevels and spread the photon frequencies over a range of 160\,kHz. Second the trigger laser has a linewidth of 50\,kHz, which is also mapped to the photons. And third, diabatically generated photons lead to an additional broadening.

Another explanation for the measured quantum-beat signal assumes photons of fixed frequency and shape, but with an emission-time jitter. In this case, we have $T_1 = \sqrt{\delta t^2 + \Delta \tau^2}$ and $T_2 = T_1 \delta t / \Delta\tau$. From these two equations, one can calculate $\delta t$, which is the lower limit of the photon duration, and $\Delta\tau$, which is the maximum emission-time jitter. In our experiment, the optimization of the single-photon source led to an increase of $\delta t$ from $0.29\,\mu$s to $0.36\,\mu$s, and at the same time to a reduction of the maximum emission-time jitter from $\Delta\tau=0.82\,\mu$s to $\Delta\tau=0.53\,\mu$s.

However, in general, both the frequency and the emission time are subject to a jitter. If  these fluctuations are uncorrelated, a whole range of $(\delta\omega, \Delta\tau)$-pairs can explain the peak and dip widths, $T_1$ and $T_2$. For our two sets of data (before (a) and after (b) the optimization of the source), this is illustrated in Figure\,\ref{fig:Parameter-Raum}. All pairs of frequency and emission-time jitters that are in agreement with the measured values of $T_1$ and $T_2$ lie on one of the two solid lines. From this figure, it is evident that the values for $\delta\omega$ and $\Delta\tau$ deduced above represent the upper limits for the respective fluctuations. Moreover, it is also nicely visible that our optimization of the source significantly improved the frequency stability and emission-time accuracy of our single-photon source.

We emphasize again that this information about the photons can be obtained from time-resolved two-photon interference experiments, but not from a measurement of the average detection probability.

\section{Conclusion}\label{sec:Conclusion}
We have shown that time-resolved two-photon interference experiments are an excellent tool to characterize single photons. In these experiments, two photons are superimposed on a beam splitter and the joint detection probability in the two output ports of the beam splitter is measured as a function of the detection-time difference of the photons. This is only possible if the photons are long compared to the detector time resolution. For identical photons, the joint detection probability is expected to be zero. Variations of the spatiotemporal modes of the photons lead to joint photodetections except for zero detection-time difference. Therefore the joint detection probability shows a pronounced dip. From the width of this dip, one can estimate the maximum emission-time jitter and the minimum coherence time of the photons. In addition, a lower limit of the single-photon duration can be obtained. This is not possible by just measuring the average detection probability with respect to the trigger producing the photons. Moreover, we have shown that a frequency difference between photons leads to a distinct oscillation in the joint detection probability. This does not only demonstrate that we are able to adjust the frequencies of the photons emitted from a single-photon source, but also that one is sensitive to very small frequency differences in time-resolved two-photon interference measurements.

\section*{Acknowledgments}
This work was supported by the Deutsche Forschungsgemeinschaft (SPP 1078 and SFB 631)  and the European Union (IST (QGATES) and IHP (CONQUEST) programs).


\bibliographystyle{elsart-harv}

\begin{thebibliography}{37}
\expandafter\ifx\csname natexlab\endcsname\relax\def\natexlab#1{#1}\fi
\expandafter\ifx\csname url\endcsname\relax
  \def\url#1{\texttt{#1}}\fi
\expandafter\ifx\csname urlprefix\endcsname\relax\def\urlprefix{URL }\fi

\bibitem[{Glauber(1965)}]{Glauber65}
Glauber, R.~J., 1965. Optical coherence and photon statistics. In: C.~DeWitt,
  A.~Blandin, C. C.-T. (Ed.), Quantum Optics and Electronics. Les Houches.
  Gordan and Breach, New York, p. 621.

\bibitem[{Gisin et~al.(2002)Gisin, Ribordy, Tittel, and Zbinden}]{Gisin02}
Gisin, N., Ribordy, G., Tittel, W., Zbinden, H., 2002. Quantum cryptography.
  Rev. Mod. Phys. 74, 145 -- 195.

\bibitem[{Knill et~al.(2001)Knill, Laflamme, and Milburn}]{Knill01}
Knill, E., Laflamme, R., Milburn, G.~J., 2001. A scheme for efficient quantum
  computing with linear optics. Nature 409, 46--52.

\bibitem[{Cabrillo et~al.(1999)Cabrillo, Cirac, Garcia-Fernandez, and
  Zoller}]{Cabrillo99}
Cabrillo, C., Cirac, J.~I., Garcia-Fernandez, P., Zoller, P., 1999. Creation of
  entangled states of distant atoms by interference. Phys. Rev. A 59,
  1025--1033.

\bibitem[{Bose et~al.(1999)Bose, Knight, Plenio, and Vedral}]{Bose99}
Bose, S., Knight, P.~L., Plenio, M.~B., Vedral, V., 1999. Proposal for
  teleportation of an atomic state via cavity decay. Phys. Rev. Lett. 83,
  5158--5161.

\bibitem[{Browne et~al.(2003)Browne, Plenio, and Huelga}]{Browne03}
Browne, D.~E., Plenio, M.~B., Huelga, S.~F., 2003. Robust creation of
  entanglement between ions in spatially seperate cavities. Phys. Rev. Lett.
  91, 067901.

\bibitem[{Duan and Kimble(2003)}]{Duan03:2}
Duan, L.-M., Kimble, H., 2003. Efficient engineering of multi-atom entanglement
  through single-photon detections. Phys. Rev. Lett. 90, 253601.

\bibitem[{Oxborrow and Sinclair(2005)}]{Oxborrow05}
Oxborrow, M., Sinclair, A.~G., 2005. Single-photon sources. Contemp. Phys. 46,
  173 -- 206.

\bibitem[{Brunel et~al.(1999)Brunel, Lounis, Tamarat, and Orrit}]{Brunel99}
Brunel, C., Lounis, B., Tamarat, P., Orrit, M., 1999. Triggered source of
  single photons based on controlled single molecule fluorescence. Phys. Rev.
  Lett. 83, 2722--2725.

\bibitem[{Lounis and Moerner(2000)}]{Lounis00}
Lounis, B., Moerner, W.~E., 2000. Single photons on demand from a single
  molecule at room temperature. Nature 407, 491--493.

\bibitem[{Moerner(2004)}]{Moerner04}
Moerner, W.~E., 2004. Single-photon sources based on single molecules in
  solids. New Journal of Physics 6, 75631--1.

\bibitem[{Darqui{\'e} et~al.(2005)Darqui{\'e}, Jones, Dingjan, Beugnon,
  Bergamini, Sortais, Messin, Browaeys, and Grangier}]{Darquie05}
Darqui{\'e}, B., Jones, M. P.~A., Dingjan, J., Beugnon, J., Bergamini, S.,
  Sortais, Y., Messin, G., Browaeys, A., Grangier, P., 2005. Controlled
  single-photon emission from a single trapped two-level atom. Science 309, 454
  -- 456.

\bibitem[{Blinov et~al.(2004)Blinov, Moehring, Duan, and Monroe}]{Blinov04}
Blinov, B.~B., Moehring, D.~L., Duan, L.~M., Monroe, C., 2004. Observation of
  entanglement between a single trapped atom and a single photon. Nature 428,
  153 -- 157.

\bibitem[{Kurtsiefer et~al.(2000)Kurtsiefer, Mayer, Zarda, and
  Weinfurter}]{Kurtsiefer00}
Kurtsiefer, C., Mayer, S., Zarda, P., Weinfurter, H., 2000. Stable solid-state
  source of single photons. Phys. Rev. Lett. 85, 290--293.

\bibitem[{Brouri et~al.(2000)Brouri, Beveratos, Poizat, and
  Grangier}]{Brouri00}
Brouri, R., Beveratos, A., Poizat, J.-P., Grangier, P., 2000. Photon
  antibunching in the fluorescence of individual color centers in diamond. Opt.
  Lett. 25, 1294--1296.

\bibitem[{Gaebel et~al.(2004)Gaebel, Popa, Gruber, Domham, Jelezko, and
  Wachtrup}]{Gaebel04}
Gaebel, T., Popa, I., Gruber, A., Domham, M., Jelezko, F., Wachtrup, J., 2004.
  Stable single-photon source in the near infrared. New Journal of Physics 6,
  77078--0.

\bibitem[{Santori et~al.(2001)Santori, Pelton, Solomon, Dale, and
  Yamamoto}]{Santori01}
Santori, C., Pelton, M., Solomon, G., Dale, Y., Yamamoto, Y., 2001. Triggered
  single photons from a quantum dot. Phys. Rev. Lett. 86, 1502--1505.

\bibitem[{Yuan et~al.(2002)Yuan, Kardynal, Stevenson, Shields, Lobo, Cooper,
  Beattie, Ritchie, and Pepper}]{Yuan02}
Yuan, Z., Kardynal, B.~E., Stevenson, R.~M., Shields, A.~J., Lobo, C.~J.,
  Cooper, K., Beattie, N.~S., Ritchie, D.~A., Pepper, M., 2002. Electrically
  driven single-photon source. Science 295, 102--105.

\bibitem[{Pelton et~al.(2002)Pelton, Santori, Vu\v{c}kovi\'{c}, Zhang, Solomon,
  Plant, and Yamamoto}]{Pelton02}
Pelton, M., Santori, C., Vu\v{c}kovi\'{c}, J., Zhang, B., Solomon, G.~S.,
  Plant, J., Yamamoto, Y., 2002. An efficient source of single photons: A
  single quantum dot in a micropost microcavity. Phys. Rev. Lett. 89, 233602.

\bibitem[{Aichele et~al.(2004)Aichele, Zwiller, and Benson}]{Aichele04}
Aichele, T., Zwiller, V., Benson, O., 2004. Visible single-photon generation
  from semiconductor quantum dots. New Journal of Physics 6, 75900--5.

\bibitem[{Hennrich et~al.(2004)Hennrich, Legero, Kuhn, and Rempe}]{Hennrich04}
Hennrich, M., Legero, T., Kuhn, A., Rempe, G., 2004. Photon statistics of a
  non-stationary periodically driven single-photon source. New Journal of
  Physics 6.

\bibitem[{Kuhn et~al.(2002)Kuhn, Hennrich, and Rempe}]{Kuhn02:2}
Kuhn, A., Hennrich, M., Rempe, G., 2002. Deterministc single-photon source for
  distributed quantum networking. Phys. Rev. Lett. 89, 067901.

\bibitem[{McKeever et~al.(2004)McKeever, Boca, Boozer, Miller, Buck, Kuzmich,
  and Kimble}]{McKeever04}
McKeever, J., Boca, A., Boozer, A.~D., Miller, R., Buck, J.~R., Kuzmich, A.,
  Kimble, H.~J., March 2004. Deterministic generation of single photons from
  one atom trapped in a cavity. Science 303, 1992--1994.

\bibitem[{Keller et~al.(2004)Keller, Lange, Hayasaka, Lange, and
  Walther}]{Keller04}
Keller, M., Lange, B., Hayasaka, K., Lange, W., Walther, H., 2004. Continuous
  generation of single photons with controlled waveform in an ion-trap cavity
  system. Nature 431, 1075--1078.

\bibitem[{{Hanbury Brown} and Twiss(1957)}]{HanburyBrown57}
{Hanbury Brown}, R., Twiss, R.~Q., 1957. Interferometry of the intensity
  fluctuations in light ii. an experimental test of the theory for partially
  coherent light. Proc. Roy. Soc. A242, 300--319.

\bibitem[{Hong et~al.(1987)Hong, Ou, and Mandel}]{Hong87}
Hong, C.~K., Ou, Z.~Y., Mandel, L., 1987. Measurement of subpicosecond time
  intervals between two photons by interference. Phys. Rev. Lett. 59,
  2044--2046.

\bibitem[{Santori et~al.(2002)Santori, Fattal, Vu\u{c}kovi\'{c}, Solomon, and
  Yamamoto}]{Santori02}
Santori, C., Fattal, D., Vu\v{c}kovi\'{c}, J., Solomon, G.~S., Yamamoto, Y.,
  2002. Indistinguishable photons from a single-photon device. Nature 419,
  594--597.

\bibitem[{Jelezko et~al.(2003)Jelezko, Volkmer, Popa, Rebane, and
  Wrachtrup}]{Jelezko03}
Jelezko, F., Volkmer, A., Popa, I., Rebane, K.~K., Wrachtrup, J., 2003.
  Coherence length of photons from a single quantum system. Phys. Rev. A 67,
  041802.

\bibitem[{Legero et~al.(2003)Legero, Wilk, Kuhn, and Rempe}]{Legero03}
Legero, T., Wilk, T., Kuhn, A., Rempe, G., 2003. Time-resolved two-photon
  quantum interference. Appl. Phys. B 77, 797.

\bibitem[{Legero et~al.(2004)Legero, Wilk, Hennrich, Rempe, and
  Kuhn}]{Legero04}
Legero, T., Wilk, T., Hennrich, M., Rempe, G., Kuhn, A., 2004. {{Quantum Beat
  of Two Single Photons}}. Phys. Rev. Lett. 93, 070503.

\bibitem[{Meystre and {Sargent III}(1998)}]{Meystre98}
Meystre, P., {Sargent III}, M., 1998. Elements of quantum optics, 3rd Edition.
  Springer-Verlag.

\bibitem[{Blow et~al.(1990)Blow, Loudon, and Phoenix}]{Blow90}
Blow, K.~J., Loudon, R., Phoenix, S. J.~D., 1990. Continuum fields in quantum
  optics. Phys. Rev. A 42, 4102--4114.

\bibitem[{Alley and Shih(1986)}]{Alley86}
Alley, C.~O., Shih, Y.~H., 1986. A new type of {EPR} experiment. In:
  Proceedings of the Second International Symposium on Foundations of Quantum
  Mechanics in the Light of New Technology, edited by M. Namiki et al., Physical
  Society of Japan, Tokyo, 47--52.

\bibitem[{Leonhardt(1997)}]{Leonhardt97}
Leonhardt, U., 1997. Measuring the Quantum State of Light. Cambridge University
  Press.

\bibitem[{Campos et~al.(1989)Campos, Saleh, and Teich}]{Campos89}
Campos, R.~A., Saleh, B. E.~A., Teich, M.~C., 1989. Quantum-mechanical lossless
  beam splitter: Su(2) symmetry and photon statistics. Phys. Rev. A 40, 1371.

\bibitem[{Leonhardt(2003)}]{Leonhardt03}
Leonhardt, U., 2003. Quantum physics of simple optical instruments. Rep. Prog.
  Phys. 66, 1207--1249.

\bibitem[{Legero(2005)}]{Legero05}
Legero, T., January 2005. {{Zeitaufgel{\"o}ste Zwei-Photonen-Interferenz}}.
  Ph.D. thesis, Technische Universit{\"a}t M{\"u}nchen.

\end{thebibliography}

\end{document}